\documentstyle[11pt,epsfig]{article}

\def\laq{\raise 0.4ex\hbox{$<$}\kern -0.8em\lower 0.62
ex\hbox{$\sim$}}
\def\gaq{\raise 0.4ex\hbox{$>$}\kern -0.7em\lower 0.62
ex\hbox{$\sim$}}

\def\vk{\vec{k}}
\def\vp{\vec{p}}
\def\vx{\vec{x}}

\begin{document}

\begin{titlepage}
\begin{flushright}
CERN-PH-TH/2005-262
\end{flushright}

\vspace{0.8cm}

\begin{center}

\huge{Dynamical back-reaction of relic gravitons}

\vspace{0.8cm}

\large{Massimo Giovannini \footnote{e-mail address: massimo.giovannini@cern.ch}}

\normalsize

\normalsize
\vspace{0.3cm}
{{\sl Centro ``Enrico Fermi", Compendio del Viminale, Via 
Panisperna 89/A, 00184 Rome, Italy}}\\
\vspace{0.3cm}
{{\sl Department of Physics, Theory Division, CERN, 1211 Geneva 23, Switzerland}}
\vspace*{2cm}

\begin{abstract}
\noindent
The dynamical  effects of the tensor modes of the geometry are investigated in the context of curvature bounces. Since the bouncing behaviour implies sharp deviations from a radiation-dominated evolution, significant back-reaction effects of relic gravitons may be expected at short wavelengths. 
 After developing a  general iterative  framework  for the calculation of dynamical back-reaction effects, explicit analytical and numerical examples are investigated for different 
 parametrizations of the energy-momentum pseudo-tensor(s) of the produced gravitons.  The reported results suggest that dynamical  back-reaction effects are a necessary ingredient for a consistent description of bouncing models 
 at late times.
\end{abstract}
\end{center}

\end{titlepage}
\newpage
\renewcommand{\theequation}{1.\arabic{equation}}
\section{Introduction}
\setcounter{equation}{0}

Relic gravitons are copiously produced in the 
early Universe due to the pumping action of the background geometry 
\cite{gris1,gris2}. If a quasi-de Sitter 
phase of expansion  is followed by a radiation-dominated 
phase, the logarithmic energy spectrum (in units of the critical  energy density), 
customarily denoted by  $\Omega_{\rm gw}(\nu)$,
is quasi-flat  \cite{star} (see also \cite{inflsp,GG1,sahni})
for present (physical) frequencies $\nu$ ranging between 
$10^{-16} \,\,{\rm Hz}$ and, approximately, $100 \,\,{\rm MHz}$. 
The transition 
from the radiation-dominated to the matter epoch produces 
an infrared branch where $\Omega_{\rm gw} \sim \nu^{-2}$ 
between $10^{-18} {\rm Hz}$ and $10^{-16} {\rm Hz}$ \cite{rub,FP,ab}.

In pre-big bang models \cite{SFD,pbb}, 
the spectrum of relic gravitons is 
far from scale-invariant. The spectral slope of 
$\Omega_{\rm gw}(\nu)$ (for frequencies larger 
than $10^{-16}$ Hz) is violet, i.e. $ \Omega_{\rm gw} \sim \nu^{\gamma}$ with $\gamma > 1$ \cite{GG1,GG3} (see also \cite{occhionero} and, in 
a complementary perspective, \cite{norma}). 
Minimal pre-big bang models are characterized by a slope 
$\gamma = 3$ that become progressively less steep 
as the frequency increases in the kHz and MHz region.

Depending on the features of the cosmological model, the energy density of relic gravitons may affect the dynamics and change 
the time evolution of the scale factor as well as of the other 
homogeneous quantities.  
If the equation of state of the sources of the 
background geometry is stiffer than radiation (i.e. $ p = w\rho$ with 
$w> 1/3$) the relativistic gravitons of short wavelengths 
can change the dynamical evolution, as it was 
noticed in the context of the stiff model of Zeldovich \cite{gris2}. 

When a stiff phase follows an inflationary stage 
of expansion \cite{mg1} (see also \cite{zelsta,fordpp}\footnote{It is 
relevant to remark that the authors of Refs.  \cite{zelsta} and \cite{fordpp}
choose to compute the energy and pressure density of the produced particles 
 by means of a perturbative expansion whose small parameter is 
the deviation from conformal coupling ($|\xi -1/6|$ in the notations 
of \cite{fordpp}). This approach (also discussed in \cite{birrel,BD}) 
may be applied, with some caveats, in the minimally coupled case 
(corresponding to $\xi \to 0$) where, it could be argued
that the expansion parameter does not exceed $1$. 
In the present investigation this expansion will not be used  since the coupling 
of relic gravitons to the background geometry is not close to conformal.
Furthermore the expansion in   $|\xi -1/6|$ may partially break down 
during the bouncing regime. When needed, numerical techniques will be employed.} ) the back-reaction 
effects of the produced gravitons set a limit on the possible 
duration of the stiff (post-inflationary) phase. 
This aspect is relevant in different situations and, for instance,
in the context of quintessential inflationary models \cite{pv1} where 
$\Omega_{\rm gw} \propto \nu$ for frequencies larger than the 
${\rm m Hz}$ \cite{mg2,mg3,sas} (see also \cite{maxdan1,li,cruise,picasso} for possible detection strategies). Blue spectral  slopes 
have been also derived in the context of quintessential models 
on the brane \cite{var1,var2} (see also \cite{sami1}). Constraints arising 
from the production of relic gravitons in quintessential models have been discussed in \cite{orito}.

Defining as $\tau$ the conformal time coordinate and $k$ as the comoving wave-number, two physically different regimes appear naturally in the problem. If $ k\tau \ll 1$ (short wavelength limit) the Fourier modes of the tensor 
fluctuations of the geometry are said to be super-adiabatically 
amplified. In the opposite regime, i.e. $k\tau \gg 1 $ (long 
wavelength limit) the tensor modes are oscillating.
In the short wavelength limit  the  effective equation of state obeyed by the relic gravitons is 
radiative, i.e. $\langle p_{\rm gw} \rangle = \langle \rho_{\rm gw}
\rangle /3$ where $\langle p_{\rm gw} \rangle$ and $\langle  \rho_{\rm gw} \rangle$
correspond to the averaged pressure and energy densities. In this case, $\langle \rho_{\rm gw} \rangle \sim a^{-4}$, where $a(\tau)$ is the scale factor.
In the long wavelength limit the effective equation of state 
may depend on the salient features of the background evolution. 
For instance, during the transition from a quasi-de Sitter stage of expansion 
to a radiation dominated stage,  the modes of long wavelengths
are compatible with an effective equation of state 
 $\langle p_{\rm gw} \rangle =- \langle \rho_{\rm gw}
\rangle /3$ implying that $ \langle \rho_{\rm gw} \rangle \sim a^{-2}$.
In this limit the system behaves, effectively, as a  generalized fluid 
 dominated by the spatial gradients of the tensor modes of the geometry.

While the conclusions mentioned in the previous 
paragraph must rely on specific definitions of the energy and pressure densities of the relic gravitons,
it is  well known that, in general, 
it is impossible to assign a localized energy density 
to the gravitational field \cite{landau}.  This caveat does not exclude 
the possibility of adopting consistent frameworks  
for the analysis of a  gravitational energy-momentum pseudo-tensor.

To be more specific, the tensor modes of the geometry 
can be characterized by a rank-two tensor $h_{ij}$ defined in the three spatial dimensions (that will taken to be flat), i.e. 
\begin{equation}
ds^2 = a^2(\tau) [ d\tau^2 - (\delta_{ij} + h_{ij}) dx^{i} dx^{j}], \qquad h_{i}^{i} = \partial_{i}h^{i}_{j} =0,
\label{pertmet}
\end{equation}
where $\delta_{ij}$ is Kroeneker symbol and $h_{ij}$, being traceless and 
divergence-less, carries two independent degrees of freedom that 
correspond to the two polarizations of a gravitational wave in a 
conformally flat Friedmann-Robertson-Walker (FRW) background \footnote{Since $h_{ij}$ is transverse and traceless,
the direction of propagation can be chosen to lie along the 
third axis and, in this case the two physical polarizations
of the graviton will be $h_{1}^{1} = - h_{2}^{2} = h_{\oplus}$
and $ h_{1}^{2} = h_{2}^{1} = h_{\otimes}$. This will be the nomenclature 
followed in the present paper.}.
The simplest way of estimating the impact of the created 
gravitons on the background dynamics is by computing the lowest-order nonlinear corrections to the Einstein tensor 
\begin{equation}
{\cal G}_{\mu}^{\nu} = R_{\mu}^{\nu} - \frac{1}{2} \delta_{\mu}^{\nu} R,
\label{ein}
\end{equation}
where $R_{\mu}^{\nu}$ and $R$ are, respectively, the Ricci tensor and the 
Ricci scalar.
The nonlinear corrections to the Einstein tensor, will consist, to lowest order,  of quadratic combinations of $h_{ij}$ that can be formally expressed as \footnote{In the 
present notations the Planck length will be defined as 
 $\ell_{\rm P} = \sqrt{8\pi G}$ 
in units $\hbar= c =1$. Natural gravitational units $16\pi G = 2 \ell_{\rm p}^2 = 1$ will also be adopted when needed.} 
\begin{equation}
\ell_{\rm P}^2 {\cal T}_{\mu}^{\nu} = - \delta^{(2)}_{\rm t} {\cal G}_{\mu}^{\nu},
\label{ps1}
\end{equation}
where the superscript at the right hand side denotes the second-order 
fluctuation of the corresponding quantity while 
the subscript refers to the tensor nature of the fluctuations. 
This procedure is essentially the one described in 
\cite{isacson1,isacson2} and has been re-explored, in a 
related context, by the authors of Refs. \cite{abramo1,abramo2} 
mainly in connection with conventional inflationary models 
where the Universe is always expanding\footnote{Recently in \cite{bab}
the field theoretical formulation of General Relativity was further 
developed with the purpose of deriving an energy-momentum tensor 
for the gravitational field. Our approach, in the present context, is more 
modest since we simply want to compare the dynamical consequences 
of different possible definitions of the energy-momentum pseudo-tensor 
of the relic gravitons.}

In \cite{ford1} (see also \cite{ford2}) a complementary perspective was 
invoked and  it was observed that, 
by perturbing the Einstein-Hilbert action to second 
order in the amplitude of the tensor modes 
of the geometry in a Friedmann-Robertson-Walker background, each single polarization 
of the graviton obeys, up to total derivatives, the action 
of a minimally coupled scalar degree of freedom
\begin{equation}
\delta_{\rm t}^{(2)} S = \frac{1}{2} \int d^{4} x\,\, a^2(\tau)\,\,\eta^{\alpha\beta} 
\partial_{\alpha} h\partial_{\beta} h,
\label{FPdef}
\end{equation}
where the dimensionless amplitude $h$ has been defined as 
\begin{equation}
h = \frac{h_{\oplus}}{\sqrt{2} \ell_{\rm P}} =\frac{h_{\otimes}}{\sqrt{2} \ell_{\rm P}}.
\label{def1}
\end{equation}
 Consequently, following \cite{ford1}, a natural ansatz for describing the back-reaction of the created 
 gravitons on the background space-time would be to use, for each single 
 polarization, the energy-momentum tensor of a minimally coupled 
 scalar field.
This approach has been exploited, for instance, in \cite{sahni} and in 
\cite{fordpp} for estimating 
the energy and pressure densities of the relic gravitons in a multi-stage
Universe. A possible 
weakness of this second approach is that Eq. (\ref{FPdef}) holds, strictly 
speaking, in a Friedmann-Roberston-Walker background.
In spite of this, it is clearly useful to bear in mind this possibility.

The purpose of the present paper is to scrutinize 
 the dynamical back-reaction effects of the relic gravitons in the extreme situation where the background geometry undergoes a contraction that smoothly evolves, through a bounce, into an expanding phase.  This 
 sequence of events takes place, for instance, in pre big-bang models.
The bouncing beahaviour is related to strong deviations 
from a radiative Universe. Consequently, 
 short wave fluctuations are expected to modify or even 
restrict the dynamical range of bouncing models since, in this 
regime, the effective equation of state of relic gravitons is the one 
of radiation. For long wavelength fluctuations, 
 the qualitative expectation  is more difficult to formulate since, as 
 already mentioned, the effective equation of state is sensitive to the specific features of the background evolution before and after the bounce.

Consider then, as an example, one of the simplest realizations of bouncing  dynamics, i.e. the one provided by minimal pre-big bang models where, in the Einstein 
frame, the Universe first contracts 
as $a(\tau) \sim \sqrt{-\tau/\tau_{1}}$ and then expands, in the 
post-big bang phase, as $a(\tau) \sim \sqrt{\tau/\tau_{1}}$. 
An accurate description of the bouncing regime 
 in the region $ -\tau_{1} <\tau <\tau_{1}$
may be achieved, for instance,  by means of a non-local potentials
\cite{mv1,mv2,gm} that 
depend on the so-called shifted dilaton which is usually introduced
 in the context of the scale-factor-duality symmetry \cite{SFD} (or more generally,
 in the treatment of $O(d,d)$ transformations \cite{jna1,jna2} acting on the background fields of the low-energy string effective action). 
 The features of the regular bouncing 
solutions discussed in \cite{maxb1,maxb2}  are such that the r\^ole of the dilaton potential 
is only crucial in the bouncing region while, away from the bounce
(i.e. $|\tau| > \tau_{1}$) the the geometry 
is driven by the kinetic energy of the dilaton field.

As soon as the tensor modes reach into the super-adiabatic regime, i.e. 
$k \tau \ll 1$ the effective 
pressure density of the relic gravitons becomes $\langle p_{\rm gw} \rangle \simeq \langle \rho_{\rm gw} \rangle$. 
Consequently, at least up to the bounce, the energy 
density of the relic gravitons  is not likely dominate on the energy density 
of the dilatonic sources since the two components evolve exactly in the 
same way with the scale factor.  
While it is plausible that only around  
the bounce the short wavelength modes become dynamically 
relevant, it is also clear that the whole 
qualitative picture should be corroborated 
by a detailed numerical and analytical treatment. 

It will be instructive to conduct the  calculation of the dynamical back-reaction within different ansatz for the energy-momentum pseudo-tensor. Then the results will be compared. A byproduct of the present analysis 
will indeed be that different forms 
of the energy-momentum pseudo-tensor lead, quantitatively, to the same 
back-reaction effects. 

The present paper is then organized as follows. In Section 2 
a class of string inspired bouncing cosmologies 
will be introduced. Section 3 is devoted to the analysis 
of the production of relic gravitons and to the 
different definitions of their energy-momentum 
pseudo-tensor in the framework of bouncing solutions. 
Section 4 deals with the analytical estimates of the effective 
barotropic indices.
Section 5 describes the implementation of a self-consistent (iterative) scheme that allows to compute the back-reaction effects of relic gravitons.  
Section 6 summarizes the main lessons 
to be drawn from the present analysis and contains also the concluding 
remarks. Finally, various technical results that are relevant for the present investigation are reported in the appendix.

\renewcommand{\theequation}{2.\arabic{equation}}
\section{A class of regular curvature bounces}
\setcounter{equation}{0}
Regular bouncing solutions can be obtained by means of   
dilaton potentials that respect the scale factor duality 
symmetry \cite{SFD}. These potentials do not depend solely upon the 
dilaton field $\phi$ but rather upon the so-called 
shifted dilaton, i.e. $\overline{\phi} = \phi - \log{\sqrt{- G_{s}}}$ 
\cite{mv1,mv2,gm} (see also \cite{oldrev}) where $G_{s}$ 
is the determinant of the four-dimensional metric in the string frame and in the cosmic time coordinate.

Therefore the starting point of the present analysis is a generally 
covariant action supplemented by a non-local dilaton potential 
 \cite{maxb1,maxb2,maxb3}.
The problem will be directly analyzed in the Einstein frame metric where 
the dilaton field and the Einstein-Hilbert term are decoupled and 
the relevant action is  \cite{maxb2}
\begin{equation}
S = \int d^{4} x \sqrt{-g} \biggl[ - \frac{R}{2 \ell_{\rm P}^2} + \frac{1}{2} 
g^{\alpha\beta} \partial_{\alpha} \phi \partial_{\beta} \phi - W(\overline{\phi})\biggr],
\label{actionB}
\end{equation}
where 
\begin{eqnarray}
&& W(\overline{\phi}) = e^{\phi} V(e^{- \overline{\phi}}),
\nonumber\\
&& e^{- \overline{\phi}} = e^{\phi/2}\int d^4 y \biggl(\sqrt{- g}\sqrt{g^{\alpha\beta}
\partial_{\alpha}\phi \partial_{\beta} \phi}\biggr)_{y} \delta(\phi(x) - \phi(y)).
\end{eqnarray}
In the following the prime denotes a derivation of the corresponding function 
with respect to his own argument (so, for instance, $V'$ will denote a derivation of $V$ with respect to $e^{-\overline{\phi}}$ and so on). 

The evolution equations derived from the action (\ref{actionB}) 
 are then \cite{maxb2}
\begin{eqnarray}
&& {\cal G}_{\mu}^{\nu} = \ell_{\rm P}^2 [T_{\mu}^{\nu}(\phi) + \tilde{T}_{\mu}^{\nu}(\phi,g)],
\label{EQ1}\\
&&g^{\alpha\beta} \nabla_{\alpha}\nabla_{\beta} \phi + e^{\phi}\biggl[ 
V - \frac{1}{2} \frac{\partial V}{\partial\overline{\phi}} - 
e^{\phi/2}\biggl(\frac{\gamma^{\mu\nu}\nabla_{\mu}\nabla_{\nu}\phi}{\sqrt{g^{\alpha\beta}
\partial_{\alpha}\phi \partial_{\beta}\phi}} {\cal I}_{1} - V' {\cal I}_{2} \biggr)\biggr]=0,
\label{EQ2}
\end{eqnarray}
where ${\cal G}_{\mu}^{\nu}$ is the Einstein tensor already introduced 
in Eq. (\ref{ein}).
In Eq. (\ref{EQ1}) $T_{\mu}^{\nu}(\phi)$ and $\tilde{T}_{\mu}^{\nu}(\phi,g)$ are, respectively,
\begin{eqnarray}
&& T_{\mu}^{\nu}(\phi) = \partial_{\mu}\phi \,\partial^{\nu} \phi - \frac{1}{2} \delta_{\mu}^{\nu} g^{\alpha\beta}\partial_{\alpha}\phi\partial_{\beta} \phi,
\label{T1}\\
&& \tilde{T}_{\mu}^{\nu}(\phi,g) = e^{\phi}\biggl( V \delta_{\mu}^{\nu} + 
\gamma_{\mu}^{\nu} \sqrt{g^{\alpha\beta}\partial_{\alpha}\phi \partial_{\beta}\phi} \,\,e^{\phi/2} \,{\cal I}_{1} \biggr);
\label{T2}
\end{eqnarray}
where the induced metric $\gamma_{\mu\nu}$ 
\begin{equation}
\gamma_{\mu\nu} = g_{\mu\nu} - \frac{\partial_{\mu}\phi \partial_{\nu}\phi}{g^{\alpha\beta}\partial_{\alpha}\phi \partial_{\beta}\phi},
\end{equation}
has been defined.
In Eqs. (\ref{EQ2}) and (\ref{T2}) the two integrals ${\cal I}_{1}$ and ${\cal I}_{2}$ are:
\begin{eqnarray}
&& {\cal I}_{1}(x) = \frac{1}{\ell_{\rm P}^3} \int d^{4} y \biggl(\sqrt{-g} V'\biggr)_{y} 
\delta(\phi(x) -\phi(y)),
\nonumber\\
&& {\cal I}_{2}(x) = \frac{1}{\ell_{\rm P}^3} \int d^{4} y \biggl(\sqrt{-g} \sqrt{g^{\alpha\beta}
\partial_{\alpha}\phi\partial_{\beta}\phi}\biggr)_{y} \delta'(\phi(x) - \phi(y)).
\end{eqnarray}

Equations (\ref{EQ1}) and (\ref{EQ2})  can now  be written for the case of a four-dimensional background geometry characterized by a conformally flat Friedmann-Robretson-Walker metric and by a homogeneous dilaton field: 
\begin{equation}
g_{\mu\nu} = a^2(\tau)\eta_{\mu\nu},\qquad \phi= \phi(\tau).
\label{BACK}
\end{equation}
Using Eq. (\ref{BACK}), the  set of homogeneous equations stemming from Eqs. (\ref{EQ1}) and (\ref{EQ2}) can be linearly combined and the 
relevant set of evolution equations, in units 
$2 \ell_{\rm P}^2=1$, becomes
\begin{eqnarray}
&& {\cal H}^2 =  \frac{{\phi'}^2}{12} + \frac{e^{\phi} a^2}{6} V,
\label{A}\\
&&  {\cal H}' =  
 - \frac{{\phi'}^2}{6} + \frac{a^2 e^{\phi}}{6}\biggl[ V - \frac{3}{2}\frac{\partial V}{\partial \overline{\phi}}\biggr],
\label{B}\\
&& \phi'' + 2 {\cal H} \phi' =- e^{\phi} a^2 \biggl[ V - \frac{1}{2} \frac{\partial V}{\partial \overline{\phi}}\biggr] ,
\label{C}
\end{eqnarray}
where the prime will denote, from now on, a derivation with respect to the conformal time coordinate $\tau$ and ${\cal H}= a'/a$.

By summing ${\cal H}'$ (from Eq. (\ref{B}) ) to $2 {\cal H}^2$ (from Eq. (\ref{A})), the right hand side of the obtained equation 
reconstructs, up to a numerical coefficient, 
the right hand side of Eq. (\ref{C}) so that 
the dependence on the potential terms may be eliminated. The 
remaining equation becomes then particularly simple: 
\begin{equation}
\frac{d}{d \tau} \biggl[ a^2 \biggl( {\cal H} + \frac{\phi'}{2}\biggr)\biggr] =0.
\label{int1}
\end{equation}
Direct integration of Eq. (\ref{int1}) gives then:
\begin{equation}
{\cal H} + \frac{\phi'}{2} = \frac{\epsilon}{a^2},
\label{int2}
\end{equation}
where $\epsilon$ is the integration constant. 
Unfortunately, in the conformal time coordinate 
the integration cannot proceed further with analytical methods.
However, Eq. (\ref{int2}) can indeed be integrated 
analytically by defining the new time coordinate $\sigma$, i.e. 
\begin{equation}
d\sigma = \frac{d\tau}{a^2(\tau)}.
\label{sigma}
\end{equation}
Since the relation of $\tau$ to cosmic time is 
$d t = a(\tau) d\tau$, the relation of the cosmic time $t$ to the $\sigma$-parametrization is simply $ dt = a^3(\sigma) d \sigma$.  Equation (\ref{sigma}) can then be used into Eq. (\ref{int2}) and the result is  
\begin{equation}
{\cal F} + \frac{d \phi}{d\sigma} = \epsilon,
\label{int3}
\end{equation}
where ${\cal F} = (\ln{a})_{,\sigma}$ and its relation to ${\cal H}$ is 
simply ${\cal H} = {\cal F}/a^2$.
By integrating once  Eq. (\ref{int3}) and by naming as $\sigma_{1}$ the 
new integration constant the following explicit relation between 
the scale factor and the dilaton field can be obtained
\begin{equation}
a(\sigma) \,\,e^{\phi(\sigma)/2} = \,e^{\epsilon (\sigma+\sigma_1)}.
\label{int4}
\end{equation}
Equation  (\ref{int4}) simply fixes a relationship between $a(\sigma)$ and 
$\phi(\sigma)$; such a relationship must hold for any potential.
Conversely, the particular form of the dilaton potential plays 
a crucial r\^ole in determining  
the relative evolution of the scale factor and of the 
dilaton whose combination appears directly in Eq. (\ref{int4}). 
Finally, once the potential is chosen,  the Hamiltonian constraint of Eq. (\ref{A})
 selects the specific value of $\epsilon$.

To proceed even further it is necessary to write Eqs.  (\ref{A}), (\ref{B}) 
and (\ref{C}) in the parametrization defined by Eq. (\ref{sigma}); in doing so  
it is useful to notice that
\begin{equation}
{\cal H} \to \frac{\cal F}{a^2},\qquad {\cal H}' \to \frac{1}{a^2}\biggl(\frac{d {\cal F}}{d\sigma} - 2 {\cal F}^2\biggr).
\label{HtoF}
\end{equation}
Consequently, Eqs. (\ref{A}), (\ref{B}) and (\ref{C}) can be written as 
\begin{eqnarray}
&&{\cal F}^2 = \frac{\xi^2}{12} + \frac{e^{\phi} a^6}{6} V.
\label{A1}\\
&& \frac{d {\cal F}}{d\sigma} =
\frac{e^{\phi} a^6}{2}\biggl[ V - \frac{1}{2} \frac{\partial V}{\partial \overline{\phi}}\biggr],
\label{B1}\\
&&\frac{d \xi }{d\sigma} = - e^{\phi} a^6 \biggl[ V - \frac{1}{2} \frac{\partial V}{\partial\overline{\phi}}\biggr],
\label{C1}
\end{eqnarray}
where $\xi = d\phi/d\sigma$. Equations (\ref{A1}) and (\ref{C1}) stem 
directly from Eq. (\ref{A}) and (\ref{C})  by using Eqs. (\ref{HtoF}). Equation (\ref{B1}) follows from Eq. (\ref{B}) by eliminating, in the obtained expression, the dependence on ${\cal F}^2$ through  Eq. (\ref{A1}). 

The solutions of the system formed by Eqs. (\ref{A1}), 
(\ref{B1}) and (\ref{C1}) can now be derived for the general class of potentials 
$V = - V_0 e^{\alpha\overline{\phi}}$. In this case we have  
\begin{eqnarray}
&& a^{6} e^{\phi}V = - V_0\,a^{6}\, e^{\phi +\alpha\overline{\phi}}= - V_0 \frac{e^{- (\alpha -2)\phi/2}}{a^{ 3(\alpha - 2)}}, 
\nonumber\\
&& a^6 e^{\phi}  \biggl[ V - \frac{1}{2} \frac{\partial V}{\partial\overline{\phi}}\biggr] =
\frac{\alpha -2}{2} \, V_0\, \frac{e^{- (\alpha -2)\phi/2}}{a^{ 3(\alpha - 2)}},
\label{pot}
\end{eqnarray}
where we used that $\overline{\phi} = \phi - 3 \ln{a_{\rm s}}$, $a_s(\tau)$ being 
the scale factor in the string frame metric. Furthermore bearing in mind that 
the scale factor in  the Einstein frame is given by $a = e^{\phi/2} a_{\rm s}$, Eqs. (\ref{pot}) follow immediately from the definition of the potential 
in terms of $\overline{\phi}$.

Equations (\ref{A1}) and (\ref{B1}) can now be combined and the result is
\begin{equation}
\frac{d {\cal F}}{d\sigma} + \frac{3}{2} (\alpha -2) {\cal F}^2 = \frac{\alpha -2}{8} \xi^2.
\label{int5}
\end{equation}
On the other hand, Eq. (\ref{int2}) allows
 to express $\xi$ in terms of $\epsilon$ and ${\cal F}$, i.e. 
\begin{equation}
\frac{\xi}{2} = \epsilon - {\cal F}.
\label{int6}
\end{equation}
Eliminating $\xi$ from Eq. (\ref{int5}) through Eq. (\ref{int6}) the whole 
problem can be reduced to the following equation
\begin{equation}
\frac{d {\cal F}}{d\sigma} = \frac{\alpha -2}{2} [\epsilon^2 - 2 {\cal F}^2 - 
2 \epsilon {\cal F}],
\label{int7}
\end{equation}
that can be easily integrated with the result that ${\cal F}$ is  
\begin{equation}
 {\cal F} = - \frac{\epsilon}{2} + \frac{\sqrt{3}\,\, \epsilon}{2}
 \tanh{\biggl[\frac{\beta(\sigma + \sigma_{2})}{2}\biggr]},
 \label{Fsol}
 \end{equation}
 where $\beta = \sqrt{3} \, \epsilon\, (\alpha - 2)$ and $\sigma_2$ 
is an integration constant.
Inserting Eq. (\ref{Fsol}) into Eq. (\ref{int6}), the explicit fom of $\xi$ turns 
out tobe
\begin{equation}
 \xi = 3 \epsilon - 
\sqrt{3}\,\,\epsilon \tanh{\biggl[\frac{\beta(\sigma + \sigma_{2})}{2}\biggr]}.
\label{xisol}
\end{equation}

 Equations (\ref{Fsol}) and (\ref{xisol}) can be integrated again so that the 
 explicit functional dependence of the dilaton and of the scale factor upon 
 the time coordinate $\sigma$ becomes:
\begin{eqnarray}
&& a(\sigma) = e^{- \phi_{\rm b}/2}e^{ - \frac{\epsilon (\sigma  - 2 \sigma_1)}{2}} \biggl[ \cosh{\frac{\beta (\sigma + \sigma_2)}{2} }\biggr]^{\frac{1}{\alpha -2 }},
\label{asol}\\
&& \phi(\sigma) = \phi_{\rm b} + 3 \epsilon \sigma - 
\frac{2}{\alpha -2} \ln{\cosh{\frac{\beta(\sigma+\sigma_{2})}{2}}}. 
\label{phisol}
\end{eqnarray}
The consistency of Eqs. (\ref{asol}) and (\ref{phisol}) with the Hamiltonian constraint (\ref{A1}) demands that the value of $\epsilon$ is fixed as:
\begin{equation}
3\epsilon^2 = V_0 e^{(\alpha -2)\phi_{\rm b}} \,\, e^{ - 3 (\alpha -2) \sigma_{1}}.
\label{consist}
\end{equation}

Equation (\ref{asol}) can now be inserted into  Eq. (\ref{asol}) to obtain the asymptotic relation (i.e. for $\sigma \to \pm \infty$) between the $\sigma$ and $\tau$. This relation is logarithmic in the sense that,  up to numerical factors 
$\sigma \simeq - \ln {(- \tau)}$ 
(for $\sigma$ and $\tau$ going to $-\infty$) and $\sigma \simeq   \ln {\tau}$ 
(for $\sigma$ and $\tau$ going to $+\infty$). This is particularly appropriate 
for numerical purposes since, in practice, the $\sigma$-parametrization 
acts as a logarithmic scale in the asymptotic regions. On the contrary, the central region of the dynamics (i.e. 
$\sigma \simeq \sigma_{2}$) is described by a linear time-scale.

To describe the features of the obtained solution 
let us fix, for simplicity, $\sigma_{1} = \sigma_2 =0$. In short the basic 
feature of the solution are the following:
\begin{itemize}
\item{} for $\alpha> 2$ the solution (\ref{asol}) and (\ref{phisol}) describes
a smooth interpolation between a contracting regime (i.e. ${\cal F}<0$) 
for $\sigma<0$ and an expanding solution, i.e. ${\cal F} > 0$ valid for 
$\sigma >0$; 
\item{} for $\alpha < 2$ the dynamical sequence is reversed 
since the scale factor expands for $\sigma <0$ and contracts for $\sigma >0$, as is can be argued from the specific form of ${\cal F}$ (see Eq. (\ref{Fsol}));
\item{} for $\alpha =2$ the solution always contract and the bouncing 
behaviour is not realized.
\end{itemize}
If $\sigma_{1}$ or $\sigma_{2}$ (or both) are different from zero the solutions
are not centered around $\sigma =0$ and the same discussion applies 
for an appropriately translated coordinate system.

Concerning the third item of the above list, 
it is interesting to  notice that, in the case $\alpha = 2$, the explicit solution reads
\begin{equation}
a(\sigma) = e^{- \phi_{\rm b}/2}\,\, e^{-\epsilon\sigma/2}, \qquad \phi(\sigma) = 3\,\epsilon\,\sigma + \phi_{\rm b}, \qquad 3 \epsilon^2 = V_0,
\end{equation} 
 implying that ${\cal F} =- \epsilon/2$.
\begin{figure}
\begin{center}
\begin{tabular}{|c|c|}
      \hline
      \hbox{\epsfxsize = 7 cm  \epsffile{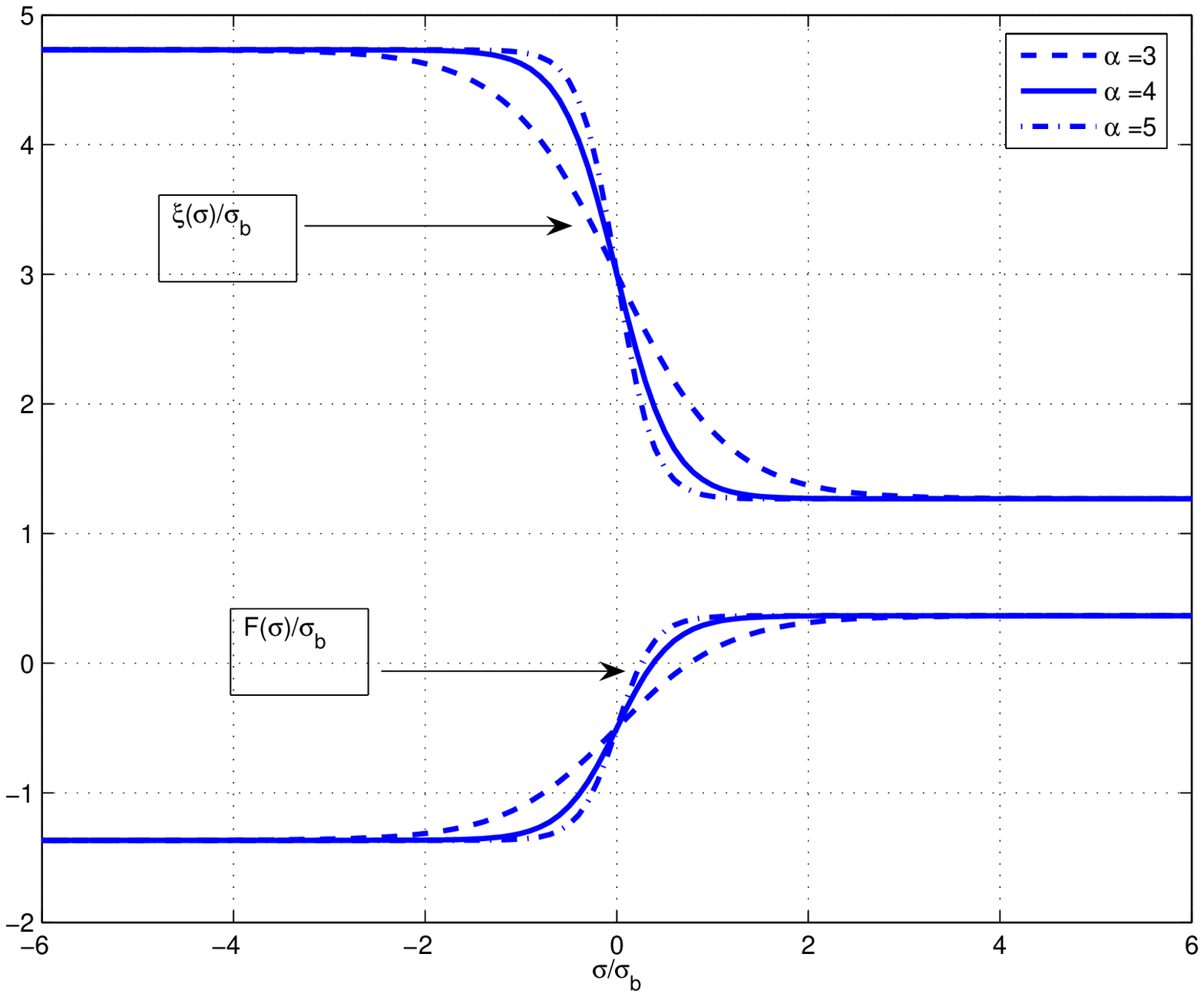}} &
      \hbox{\epsfxsize = 7.4 cm  \epsffile{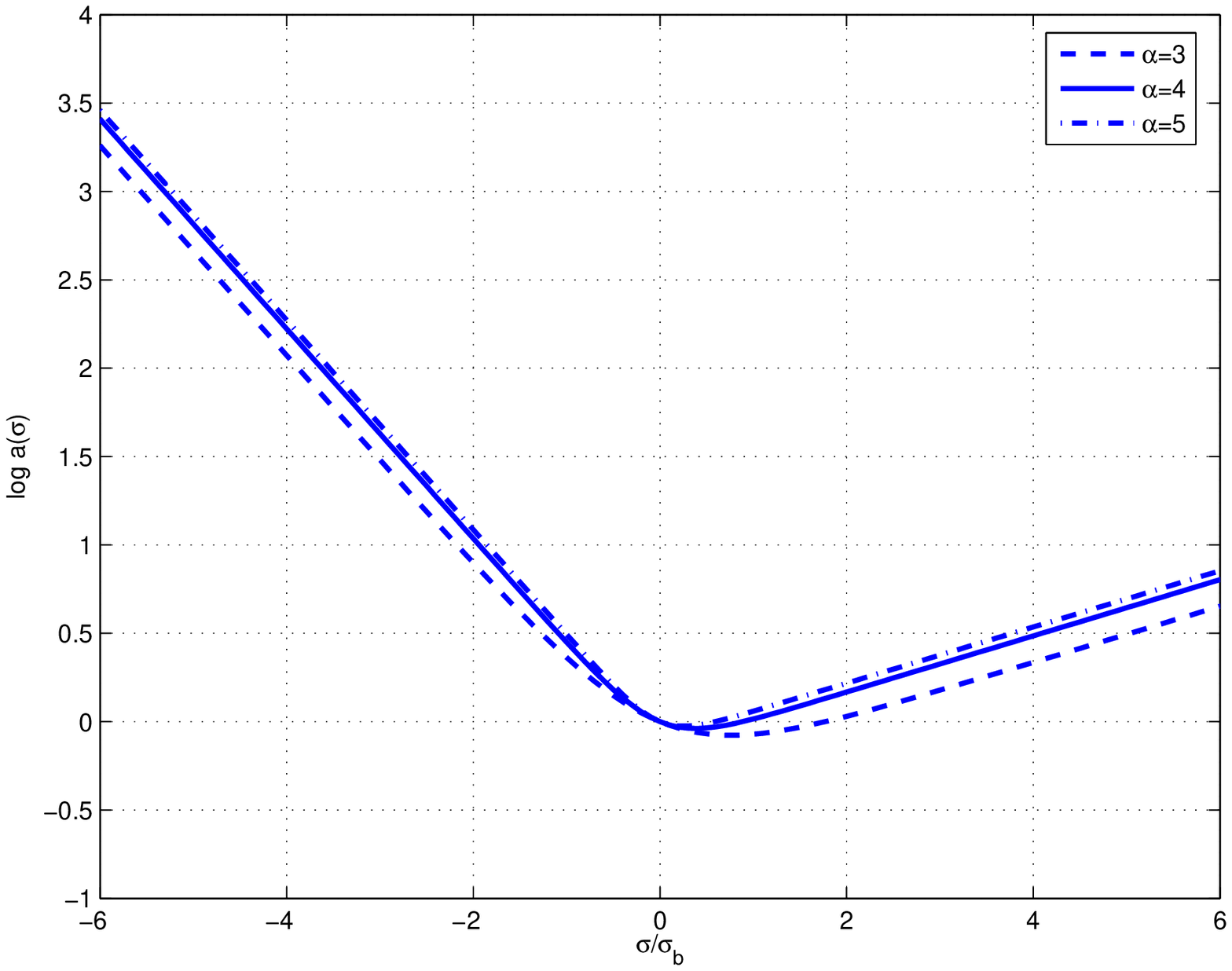}}\\
      \hline
\end{tabular}
\end{center}
\caption{In the left plot the evolution of ${\cal F}/\sigma_{\rm b}$ and 
$\xi/\sigma_{\rm b}$ are illustrated for different values of the parameter $\alpha$.
In the right plot the logarithm (to base 10) of the scale factor is illustrated 
for different values of $\alpha$.} 
\label{figure1}
\end{figure}
In Fig. \ref{figure1} (left panel) the evolution of $\xi(\sigma)$ and  ${\cal F}(\sigma)$ is
illustrated as it emerges from Eqs. (\ref{Fsol}) and (\ref{xisol}). The 
time scale is given in terms of $\sigma_{\rm b} =1/\epsilon$, i.e. 
the typical bouncing time. In the left plot, the logarithm\footnote{In the present 
paper the natural logarithm will always be denoted as $\ln$ 
while the logarithm to base $10$ will be denoted as $\log$.}
 (to base $10$) of the scale factor is illustrated for the same range of $\alpha$.
The initial conditions for the solutions can be charcterized in terms of $\phi$ or 
more precisely $e^{\phi/2}$ that measures the initial strength 
of the gauge coupling. We shall be primarily interested in the case when 
the solution contracts for $\sigma <0$ and expands for $\sigma >0$ (i.e.
$\alpha >2$). In this situation the pre-big bang initial conditions 
will be reproduced since $e^{\phi/2}\ll 1$ for $\sigma \to -\infty$.
Notice that  $\phi_{\rm b}$ measures the strength of the gauge coupling 
around the bounce, i.e. for $\sigma \simeq 0$. Of particular interest will be 
the case when $\phi_{\rm b} \simeq 0$, i.e. when strong coupling 
is reached just at the bounce. It is clearly always possible to tune 
$\phi_{\rm b}$ in such a way that the gauge coupling is much smaller 
than $1$ but we shall try to avoid this tuning and examine the 
more realistic situation where $\phi_{\rm b}\simeq 0$.

The $\sigma$-parametrization is particularly useful 
for the numerical evolution of the fluctuations and for the self-consistent 
treatment of the dynamical back-reaction effects. It is sometimes practical, 
in a complementary perspective, 
to solve for the mode functions directly in the conformal 
time parametrization. As already remarked, in the conformal time 
parametrization we cannot rely on exact solutions. Therefore 
the idea will be to map the asymptotic solutions 
in the $\sigma$ parametrization into asymptotic 
solutions in the $\tau$ parametrization and then solve numerically 
Eqs. (\ref{A}), (\ref{B}) and (\ref{C}).  It is therefore necessary to develop 
the numerical intuition for the evolution of the background 
in the conformal time parametrization. Of particular 
technical relevance is the precise mapping of the  initial conditions  from the 
$\sigma$-parametrization to the conformal time coordinate.

To fix the ideas, consider the case $\alpha =4$ with 
the bounce occurs around $\sigma_{1} = \sigma_{2} =0$. In this case 
the mapping of the initial conditions between the $\sigma$ 
and the $\tau$ parametrization can be derived from the following 
chain of relations
\begin{equation}
3 \epsilon^2 = V_0 e^{2 \phi_{\rm b}} = \frac{e^{-2\phi_{\rm b}}}{\tau_{1}^2},
\label{consist2}
\end{equation}
 where the first equality is nothing but 
 Eq. (\ref{consist}) in the case $\alpha= 4$ (and $\sigma_1=0$). The second equality is fixed by solving directly the Hamiltonian 
 constraint of Eq. (\ref{A}) in the 
 asypmptotic regions (i.e. $\tau \to \pm \infty$). Equation (\ref{consist2}) 
 fixes the relation between the bouncing time in the $\sigma$-parametrization,
 i.e. $\sigma_{\rm b}= 1/\epsilon$,
 in terms of the bouncing time in the $\tau$-parametrization, i.e. $\tau_{1}$.
\begin{figure}
\begin{center}
\begin{tabular}{|c|c|}
      \hline
      \hbox{\epsfxsize = 7 cm  \epsffile{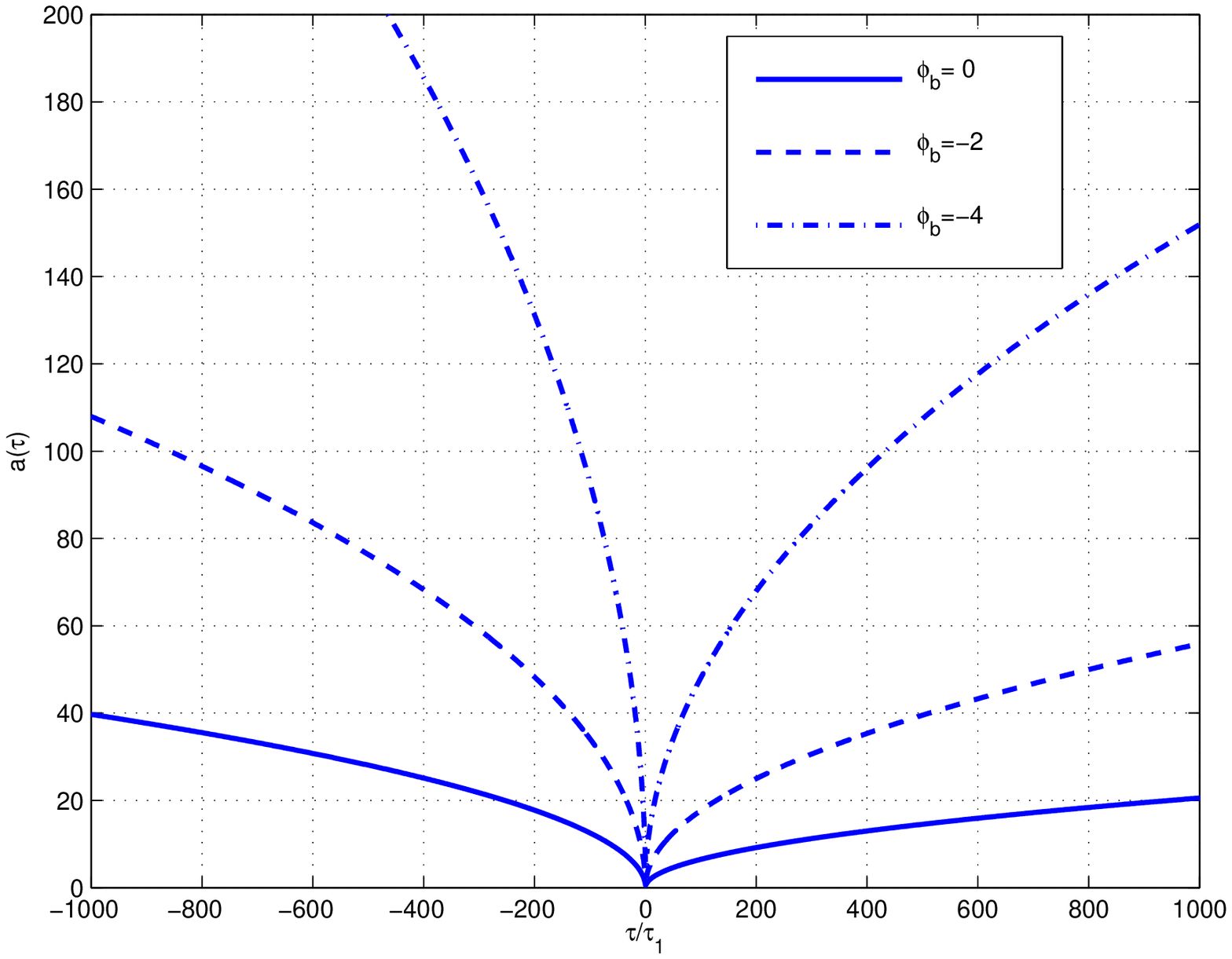}} &
      \hbox{\epsfxsize = 6.9 cm  \epsffile{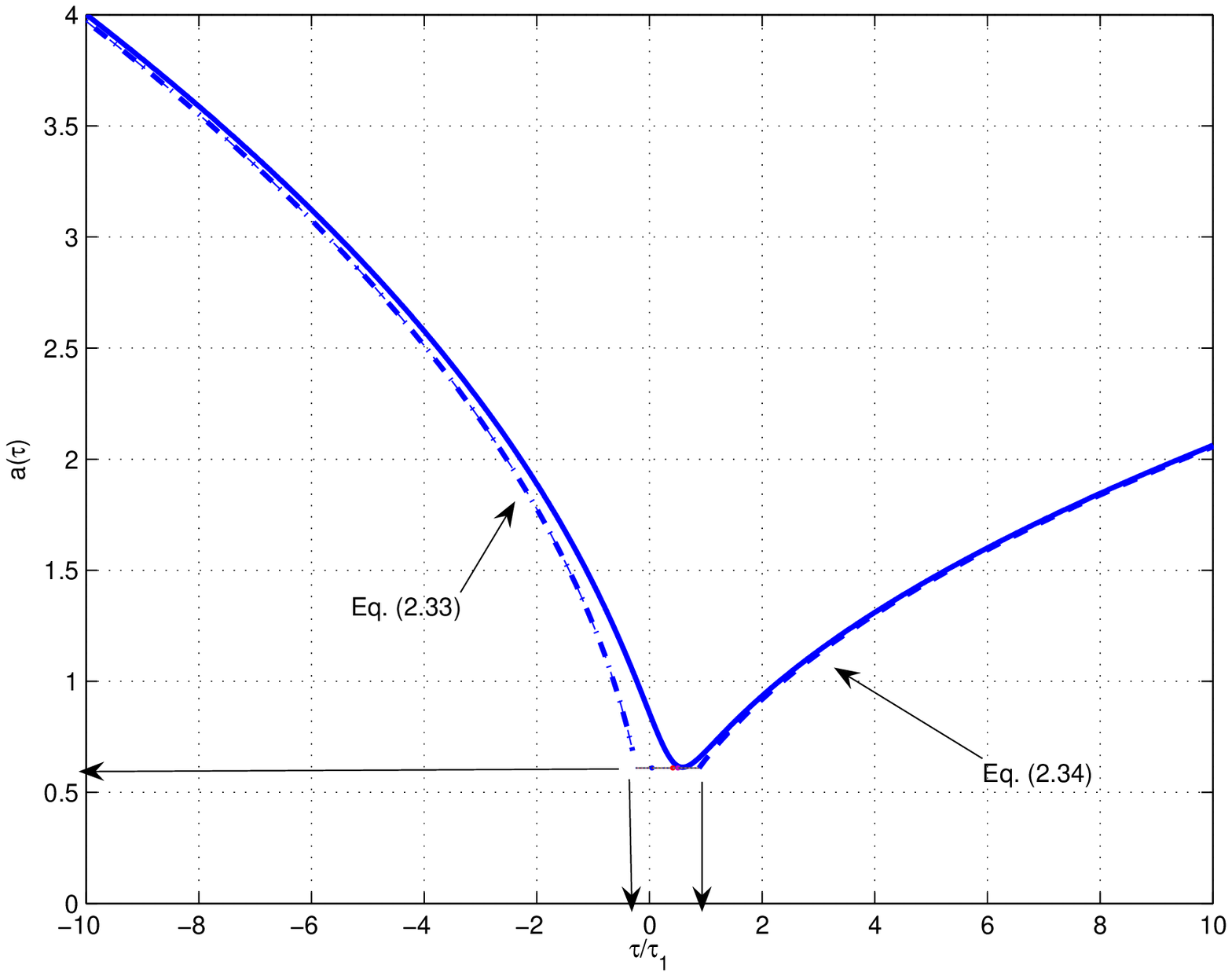}}\\
      \hline
\end{tabular}
\end{center}
\caption{ The evolution 
of the scale factor (left panel) is illustrated for different initial values of the gauge coupling. In the right plot the
analytical approximations for  the evolution of the scale factor (dashed lines)
are compared with the numerical solution.} 
\label{figure2}
\end{figure}
In Fig. \ref{figure2} the numerical integration of Eqs. (\ref{A}), (\ref{B}) and 
(\ref{C}) is illustrated in terms  of the scale factor.
In particular, for $\tau < -\tau_{1}$ the solution is  well approximated by 
\begin{equation}
a(\tau) = a_{-} \sqrt{\frac{-\tau}{2\tau_{1}}}, \qquad a_{-} = e^{- \phi_{\rm b}/2} \sqrt{\frac{2 (\sqrt{3} +1)}{\sqrt{3}}},
\label{anal1}
\end{equation}
while for $\tau> -\tau_{1}$ the solution is well approximated by 
\begin{equation}
a(\tau) = a_{+} \sqrt{\frac{\tau}{2\tau_{1}}}, \qquad a_{+} = e^{- \phi_{\rm b}/2} \sqrt{\frac{2 (\sqrt{3} -1)}{\sqrt{3}}},
\label{anal2}
\end{equation}
where $\phi_{\rm b}$ is the same constant introduced in Eq. (\ref{phisol}). 
In the pre-bounce region the evolution of the dilaton can be obtained 
from Eq. (\ref{phisol}) by using the asymptotic relation between $\tau$ 
and $\sigma$ in the limit $\tau \to -\infty$ and $\sigma\to -\infty$. The result is:
\begin{equation}
\phi(\tau) = \phi_{\rm b} - (\sqrt{3} -1)\ln{2} - \sqrt{3} \ln{\biggl(\frac{\sqrt{3} +1}{\sqrt{3}}\biggr)} 
- \sqrt{3} \ln{\biggl( - \frac{\tau}{\tau_{1}}\biggr)}.
\label{phanal}
\end{equation}

In the intermediate region, the minimum of the scale factor corresponds to 
\begin{equation}
a_{\rm min} = 0.61\, e^{-\phi_{\rm b}/2}.
\end{equation}
The ``width" of the bounce is illustrated in Fig. \ref{figure2} (right plot)
and  can be estimated analytically as 
\begin{equation}
\frac{\tau_{-}}{\tau_{1}} = - \frac{0.74}{a_{-}^2} e^{- \phi_{\rm b}},\qquad 
\frac{\tau_{+}}{\tau_{1}} =  \frac{0.74}{a_{+}^2} e^{- \phi_{\rm b}}.
\end{equation}
In Fig. \ref{figure2} the values of $\tau_{\pm}$ correspond to the two 
vertical arrows around the origin.

By changing the various parameters it appears that both $a^2 H^2$ and ${\phi'}^2/2$ 
are not sensitive to the value of $\phi_{\rm b}$ but rather they depend on $\tau_{1}$.
In particular, it turns out that 
\begin{equation}
{\cal H}^2 \simeq \frac{0.61}{\tau_{1}^2},\qquad 
\frac{{\phi'}^2}{2} \simeq \frac{7.5}{\tau_{1}^2}.
\end{equation}

In the case $\alpha =2$ the solution can be expressed exactly also in the conformal 
time parametrization. Recalling Eq. (\ref{sigma}) we obtain that 
\begin{equation}
\sigma = - \frac{1}{\epsilon} \ln{(- \epsilon)\tau},\qquad 
a(\tau) = e^{-\phi_{\rm b}/2} \sqrt{- \epsilon\tau},\qquad \phi(\tau) = \phi_{\rm b} 
- 3 \ln{(-\epsilon\tau)}.
\end{equation}
Thus, unlike the case $\alpha >3$ the solution with $\alpha =2 $ 
is singular in the conformal time parametrization.

The $\sigma$ parametrization has one more important feature that has been 
already implicitly treated in Eq. (\ref{HtoF}). While the sign of ${\cal H}$ 
and ${\cal F}$ is preserved by the mapping $\tau\to \sigma$, the sign 
of ${\cal H}'$ is, in general, different from the sign of ${\cal F}_{,\sigma}$. 
The simplest way of realizing this aspect is to notice that, from 
Eqs. (\ref{sigma}) and (\ref{HtoF}) we can write 
\begin{equation}
{\cal F}  = a^2 {\cal H},\qquad \frac{\partial {\cal F}}{\partial \sigma} = a^4( {\cal H}' + 2 {\cal H}^2).
\label{mapp1}
\end{equation}
Consider now the simple case of a radiation-dominated Universe.  In this 
case $a(\tau) = (\tau + 2 \tau_{1})/\tau_{1}$ for $\tau \geq -\tau_{1}$. 
Consequently, from Eq. (\ref{mapp1}), ${\cal H}$ and ${\cal F}$  will have the same sign but ${\cal F}_{,\sigma}$ will be always 
positive while ${\cal H}' = - 1/(\tau + 2 \tau_{1})^2$ will always be negative.
The cure for this type of problem is very simple. The invariant quantities 
(like for instance) the Ricci scalar are always the same in both coordinate systems. So, for instance, in a radiation dominated Universe 
the background  Ricci scalar, i.e. $\overline{R}$ 
 is always zero both in the $\tau$ and in the $\sigma$ parametrizations.
 In fact, 
\begin{equation}
\overline{R} = - \frac{6}{a^2} ( {\cal H}^2 + {\cal H}') = - \frac{6}{a^6}\biggl( \frac{\partial {\cal F}}{\partial \sigma} - {\cal F}^2\biggr),
\label{curv}
\end{equation}
where the second equality follows by using Eq. (\ref{HtoF}). Now, it is easy to show that, for a radiation-dominated Universe, the evolution equation
obeyed by ${\cal F}$ is exactly ${\cal F}_{,\sigma }= {\cal F}^2$. 

It is appropriate to remark at this point that the bouncing solutions 
obtained here in the framework of the low-energy string effective action 
differ from the ones derived in the presence of spatial curvature and 
in a general relativistic context. For instance in \cite{deruelle1,deruelle2} 
(see also \cite{tur,hwang} and references therein)
spatially 
closed models driven by a massive scalar were studied. As anticipated, 
in this framework the dynamical sequence of events differs from the one discussed here. In \cite{deruelle1,deruelle2}  the Universe 
first deflates, then bounces and finally inflates \footnote{This class of closed 
bounces has a long history. For instance in \cite{ellis1,ellis2} useful 
analytical solutions were derived and exploited. In connection with 
this problem see also \cite{full,star2,matzner}.}.  In spite of this statement, the treatment the dynamical back-reaction of relic gravitons in closed bounces 
can be developed by following the same methods discussed here.

\renewcommand{\theequation}{3.\arabic{equation}}
\section{Energy-momentum pseudo-tensor(s)}
\setcounter{equation}{0}
\subsection{General considerations}
Recalling Eq. (\ref{pertmet}), the quadratic action for $h_{ij}$ can be
obtained by perturbing Eq. (\ref{actionB}) to second-order in the 
amplitude of the tensor fluctuations of the metric. Some
 relevant expressions can be found in the 
 appendix A. Up to total 
derivatives, the result can be written as 
\begin{equation}
\delta_{\rm t}^{(2)} S = S_{\rm gw} = \frac{ 1}{8 \ell_{\rm P}^2} \int d^3 x\,\, d\tau\,\, \sqrt{- \overline{g}} \partial_{\alpha} 
h_{i}^{j} \partial_{\beta} h_{j}^{i} \overline{g}^{\alpha\beta},
\label{action1}
\end{equation}
where $\overline{g}_{\alpha\beta} = a^2(\tau) \eta_{\alpha\beta}$ and 
$\eta_{\alpha\beta}$ is the Minkowski metric.

As discussed in connection with  Eq. (\ref{pertmet}), the two polarizations
of the graviton can be chosen to be 
 $h_{1}^{~1}= - h_{2}^{~2} = h_{\oplus}$ 
and $h_{1}^{~2}= h_{2}^{~1} = h_{\otimes}$. Furthermore recalling the notations of  Eq. (\ref{def1}),
Eq. (\ref{action1}) becomes, for a single tensor polarization
\begin{equation}
S_{\rm gw} = \frac{1}{2} \int d^{3} x \,\, d\tau \,\, \sqrt{-\overline{g}}\,\,\partial_{\alpha} h \partial_{\beta} h \overline{g}^{\alpha\beta},
\label{TA1}
\end{equation}
Using the parametrization defined in Eq. (\ref{sigma}),  Eq. (\ref{TA1}) can be 
also written as 
\begin{equation}
S_{\rm gw} = \frac{1}{2} \int d^{3} x \,\, d\sigma \,\, \biggl[ 
\biggl(\frac{\partial h}{\partial \sigma}\biggr)^2 - a^4(\sigma) \partial_{m} h \partial^{m}h
\biggr].
\label{TA2}
\end{equation}
The associated canonical momentum, i.e.  $\Pi= \partial_{\sigma} h$ allows 
to obtain the  Hamiltonian of the tensor modes, i.e.
\begin{equation}
H_{\rm gw}(\sigma) =\frac{1}{2} \int d^3 x \biggl[ \Pi^2 + a^4 \partial_{m}h \partial^{m} h\biggr].
\label{H1t}
\end{equation}
By promoting the classical fields to quantum mechanical operators 
we have that \footnote{Notice that, within our set of conventions, 
$\hat{h}_{-\vk}^{\dagger} = \hat{h}_{\vk}$ and $\hat{\Pi}_{-\vk}^{\dagger} = \hat{\Pi}_{\vk}$.}
\begin{eqnarray}
&& \hat{h}(\vec{x},\sigma) = \frac{1}{2 ( 2\pi)^{3/2}} \int d^{3} k
 [ \hat{h}_{\vk}(\sigma) e^{ - i \vec{k} \cdot \vec{x}} + \hat{h}_{\vk}^{\dagger}(\sigma) e^{ i \vec{k} \cdot\vec{x}}],
\nonumber\\
&& \hat{\Pi}(\vec{x},\sigma) = \frac{1}{2 ( 2\pi)^{3/2}} \int d^{3} k [ \hat{\Pi}_{\vk}(\sigma) e^{ - i \vec{k} \cdot \vec{x}} + \hat{\Pi}_{\vk}^{\dagger}(\sigma) e^{ i \vec{k} \cdot\vec{x}}].
\label{oper}
\end{eqnarray}
In the Heisenberg representation the evolution of the field operators can be 
written as 
\begin{eqnarray}
\hat{h}_{\vk}(\sigma) &=& \hat{a}_{\vk}(\sigma_{\rm i}) F_{k}(\sigma) + \hat{a}_{-\vk}^{\dagger}(\sigma_{\rm i}) F_{k}(\sigma)^{*},
\label{Heis1a}\\
\hat{\Pi}_{\vk}(\sigma) &=&  \hat{a}_{\vk}(\sigma_{\rm i}) G_{\vk}(\sigma) + 
\hat{a}_{-\vk}^{\dagger}(\sigma_{\rm i}) G_{k}(\sigma)^{*}, 
\label{Heis1b}
\end{eqnarray}
where $\hat{a}_{\vk}(\sigma_{\rm i})$ and $\hat{a}_{-\vk}(\sigma_{\rm i})$ 
annihilate the vacuum 
for $\sigma_{\rm i} \to -\infty$. 
The mode functions $F_{k}$ and $G_{k}$  now obey 
\begin{equation}
\frac{d^2 F_{k}}{d\sigma^2} + \Omega^2 F_{k} =0,
\qquad G_{k} = \frac{d F_{k}}{d\sigma},
\label{modefun}
\end{equation}
where $\Omega= k a^2(\sigma)$.

\subsection{Mixing coefficients}
In the short 
wavelength limit, i.e. $k |\tau| > 1$ the mode function $F_{k}(\sigma)$ has 
an oscillating behaviour. This regime characterizes the asymptotics 
of the evolution of the evolution of the mode functions. Initially, for 
$\sigma \to -\infty$ the approximate solution of Eq. (\ref{modefun})
is simply given by 
\begin{equation}
F_{k}(\sigma) = \frac{1}{\sqrt{2 \Omega}}e^{- i\int \Omega d\sigma},\qquad
G_{k}(\sigma) = - \frac{1}{\sqrt{ 2 \Omega}} [ {\cal F} + i \Omega] e^{- i \int \Omega d\sigma}.
\label{MFI}
\end{equation} 
In the limit $\sigma \to +\infty$ the solution can be expressed by means of two 
mixing coefficients $c_{+}(k)$ and $c_{-}(k)$, i.e. 
\begin{eqnarray}
&&F_{k}(\sigma) = \frac{1}{\sqrt{ 2 \Omega}} \biggl[ c_{+}(k) e^{- i \int \Omega d\sigma} 
+ c_{-}(k) e^{i \int \Omega  d\sigma}\biggr],
\nonumber\\
&& G_{k}(\sigma)= \frac{1}{\sqrt{2 \Omega}}\biggl[ - ( {\cal F} + i \Omega)
c_{+}(k) e^{- i \int \Omega d\sigma} + ( i \Omega - {\cal F}) c_{-}(k) e^{ i \int \Omega d\sigma} \biggr].
\label{MFII}
\end{eqnarray}
In the long wavelength limit, i.e. $k|\tau| < 1$ the approximate solution of Eq. (\ref{modefun}) across the bounce is even simpler and it is given by 
\begin{equation}
F_{k} = A_{k} + B_{k} \sigma, \qquad G_{k} = B_{k}.
\label{MFIII}
\end{equation}
It is relevant to remark here that the condition $k |\tau|\simeq 1$ reads, in the 
$\sigma$-parametrization,
\begin{equation}
|\int \Omega d\sigma| \simeq 1.
\label{condOM}
\end{equation} 
Due to the presence of the absolute values, there will be 
two distinct moments, in the dynamical evolution, where 
the condition $ k |\tau | =1$ is verified. Before the bounce, i.e. for $\sigma <0$, 
for a given mode $k$ the condition $- k \tau(\sigma) \simeq 1$ will define 
the time of ``exit" in the $\sigma$ parametrization.  The notation $\tau(\sigma)$ 
simply means that the specific relation between $\tau$ and $\sigma$ has to be 
derived from Eq. (\ref{sigma}) by using the appropriate expression of the scale
factor.
After the bounce, i.e. 
for $\sigma >0$, the condition $k\tau(\sigma) \simeq 1$ will define the time 
of ``reentry" in the $\sigma$-parametrization.
 
Defining conventionally the typical time scale of the bounce
as $\sigma_{\rm b} = 1/\epsilon$,  consider the expression of the scale 
factor given in Eq. (\ref{asol}). In units of bouncing times 
the exit and reentry occur, approximately, for 
\begin{eqnarray}
\frac{\sigma_{\rm ex}}{\sigma_{\rm b}} &=&
 - \frac{1}{\sqrt{3} + 1} \ln{\biggl[ \frac{\sqrt{3} +1}{\kappa} 
 e^{\phi_{\rm b}}\, 2^{2/(\alpha -2)}\biggr]},
\label{exit}\\
\frac{\sigma_{\rm re}}{\sigma_{\rm b}} &=&
  \frac{1}{\sqrt{3} - 1} \ln{\biggl[ \frac{\sqrt{3} -1}{\kappa} 
 e^{\phi_{\rm b}}\, 2^{2/(\alpha -2)}\biggr]},
\label{reentry}
\end{eqnarray}
where we defined $\kappa = k \sigma_{\rm b}$, i.e. the wave-number in units 
of inverse bounce time.
Equations (\ref{exit}) and (\ref{reentry}) can be easily derived from Eq. (\ref{condOM})
by using the explicit expressions of $a(\sigma)$ given in Eq. (\ref{asol}) evaluated, respectively, in the limits $\sigma < -\sigma_{\rm b}$ and $\sigma > \sigma_{\rm b}$.
Equations (\ref{exit}) and (\ref{reentry}) are derived by fixing $\sigma_1 =\sigma_2 =0$ in  Eq. (\ref{asol})  and by using Eq. (\ref{sigma}).
Moreover, because of the considerations elaborated in the previous Section, 
the expressions given in Eqs. (\ref{exit}) and (\ref{reentry}) hold for $\alpha >2$.

For modes $\kappa < 1$, the three solutions given in
 Eqs. (\ref{MFI}), (\ref{MFII})
and (\ref{MFIII}) can be appropriately matched in such a way that the field 
operators are continuous function of $\sigma$. This procedure 
leads to the determination of the coefficients $A_{k}$, $B_{k}$
\begin{equation}
A_{k} = \frac{1}{\sqrt{2 \Omega_{\rm ex}}} e^{- i k\tau_{\rm ex}}[ 1 + ({\cal F}_{\rm ex} + i \Omega_{\rm ex})\sigma_{\rm ex}],\qquad
B_{k} = -  \frac{1}{\sqrt{2 \Omega_{\rm ex}}} e^{- i k\tau_{\rm ex}} ({\cal F}_{\rm ex} + i \Omega_{\rm ex}),
\label{AB}
\end{equation}
and, ultimately, of the mixing coefficients $c_{\pm}(k)$:
 \begin{eqnarray}
&& c_{-}(k) = - \frac{i}{ 2 \sqrt{\Omega_{\rm ex} \Omega_{\rm re}}} e^{- i k ( \tau_{\rm re} + \tau_{\rm ex})}[ ({\cal F}_{\rm re}  - {\cal F}_{\rm ex})
\nonumber\\
 &&+ 
i ( \Omega_{\rm re} - \Omega_{\rm ex}) - ({\cal F}_{\rm re} + i \Omega_{\rm re})({\cal F}_{\rm ex} + i \Omega_{\rm ex}) ( \sigma_{\rm re} - \sigma_{\rm ex})],
\label{CM}\\
&& c_{+}(k) = - \frac{i}{ 2 \sqrt{\Omega_{\rm ex} \Omega_{\rm re}}} e^{- i k ( \tau_{\rm ex} - \tau_{\rm re})}[ ({\cal F}_{\rm ex}  - {\cal F}_{\rm re}) 
\nonumber\\
&&+ 
i ( \Omega_{\rm re} + \Omega_{\rm ex}) + ({\cal F}_{\rm re} - i \Omega_{\rm re})({\cal F}_{\rm ex} + i \Omega_{\rm ex}) ( \sigma_{\rm re} - \sigma_{\rm ex})].
\label{CP}
\end{eqnarray}
In Eqs. (\ref{AB}), (\ref{CM}) and (\ref{CP}), the obvious notations 
$\Omega_{\rm ex} = \Omega(\sigma_{\rm ex})$, ${\cal F}_{\rm ex}= {\cal F}(\sigma_{\rm ex})$ 
(and similarly for $\Omega_{\rm re}$ and ${\cal F}_{\rm re}$) have been employed.
Equations (\ref{CM}) and (\ref{CP}) imply that 
$|c_{+}(k)|^2 - |c_{-}(k)|^2 =1$ which is a simple consequence 
of the unitary evolution.
For $\kappa > 1$ the mixing coefficient is exponentially suppressed. To 
assess accurately the form of exponential suppression, a numerical treatment 
is necessary. The strategy here is to integrate numerically Eqs. (\ref{modefun}) 
with initial conditions given by Eq. (\ref{MFI}) and to obtain, numerically 
the mixing coefficients, i.e. 
\begin{eqnarray}
|c_{+}(k)|^2 - |c_{-}(k)|^2 &=& i [F_{k}^{*}(\sigma) G_{k}(\sigma) - F_{k}(\sigma) G_{k}^{*}(\sigma) ],
\label{DIFF}\\
|c_{+}(k)|^2 + |c_{-}(k)|^2 &=& \frac{1}{\Omega} \{ |G_{k}(\sigma)|^2 +
( {\cal F}^2 +\Omega^2)  |F_{k}(\sigma)|^2
\nonumber\\
&+& {\cal F}[ F_{k}^{*}(\sigma) G_{k}(\sigma)+ F_{k}(\sigma) G_{k}^{*}(\sigma)]\}.
\label{SUM}
\end{eqnarray}
For the numerical consistency of the whole approach, it must be true for 
each specific numerical integration, that the combination reported 
in Eq. (\ref{DIFF}), corresponding to the Wronskian of the solution, 
 is equal to $1$ (since it is chosen to be equal to one at the initial 
 integration time, see  Eq. (\ref{MFI})). The numerical accuracy 
 of the Wronskian normalization condition is  a check of the 
 consistency of the whole approach. Along the same 
 lines, the combination in Eq. (\ref{SUM}) must go to a constant 
 for $\sigma\to +\infty$. 
 
A set of first-order equations valid for the mixing coefficients will now be discussed. Defining
\begin{equation}
e_{\pm} = e^{\pm i\int \Omega d\sigma},
\end{equation}
it is possible to find that the mixing coefficients obey
\begin{equation}
\frac{ d c_{+}}{d\sigma} = \frac{1}{2}\frac{d \ln \Omega}{d\sigma}\,\, e_{+}^2  c_{-},
\qquad \frac{ d c_{-}}{d\sigma} = \frac{1}{2}\frac{d \ln \Omega}{d\sigma}\,\, e_{-}^2  c_{+}.
\label{effeq}
\end{equation}
Equation (\ref{effeq}) was derived, in a related context, in Ref. \cite{zelsta}.
Since $c_{+}(k)$ and $c_{-}(k)$ are two complex quantities subjected 
to the condition $|c_{+}(k)|^2 - |c_{-}(k)|^2 =1$, they correspond, overall
to three real variables that can be chosen to be 
\begin{eqnarray}
&& \overline{n}_{k} = |c_{-}(k)|^2,
\nonumber\\
&& p_{k} = c_{+}(k) c_{-}^{*}(k) e_{-}^2 + c_{-}(k) c_{+}^{*}(k) e_{+}^2,
\nonumber\\
&& q_{k} = i [ c_{+}(k) c_{-}^{*}(k) e_{-}^2 - c_{+}^{*}(k) c_{-}(k) e_{+}^2].
\end{eqnarray}
The quantities  
$\overline{n}_{k}$ , $p_{k}$ and $q_{k}$ obey the following set of equations
\begin{eqnarray}
&& \frac{d \overline{n}_{k}}{d\sigma} = {\cal F} p_{k},
\label{ss1}\\
&& \frac{d p_{k}}{d\sigma} = 2 {\cal F} ( 1 + 2 \overline{n}_{k}) - 2 \Omega q_{k},
\label{ss2}\\
&& \frac{d q_{k}}{d\sigma} = 2 \Omega p_{k}.
\label{ss3}
\end{eqnarray}
The initial conditions should be such that, initially, $c_{+}(k) =1$ and $c_{-}=0$, i.e. 
\begin{equation}
\overline{n}_{k}(\sigma_{\rm i}) =0, \qquad p_{k}(\sigma_{\rm i}) = 0, \qquad 
q_{k}(\sigma_{\rm i}) =0.
\label{ssinc}
\end{equation}
Now the analytical estimates will be corroborated by numerical 
integration. One of the pleasant features of the $\sigma$- 
parametrization is that the spectrum of the canonical momenta 
is strictly constant in the long wavelength limit. The spectrum of canonical momenta is the Fourier 
transform of the two-point correlation function of $\hat{\Pi}(\vx, \sigma)$, i.e.
\begin{equation}
\langle \hat{\Pi}(\vec{x}, \sigma) \hat{\Pi}(\vec{y},\sigma) \rangle 
= \int d \ln{k}\,\, |\delta_{\Pi}(k,\sigma)|^2 \frac{\sin{k r}}{kr},
\label{momcor1}
\end{equation}
where 
\begin{equation}
|\delta_{\Pi}(k,\sigma)|^2 = \frac{k^3}{2\pi^2} |G_{k}(\sigma)|^2.
\label{momcor2}
\end{equation}
Equations (\ref{momcor1}) and (\ref{momcor2}) can be obtained 
by using Eqs. (\ref{oper}) together with Eqs. (\ref{Heis1a}) and (\ref{Heis1b}). When performing the appropriate 
quantum mechanical expectation values (in the Heisenberg representation) 
it should be recalled that 
$\langle \hat{a}_{\vk} \hat{a}^{\dagger}_{\vp}\rangle 
= \delta^{(3)}(\vec{k} - \vec{p})$.
 
\begin{figure}
\begin{center}
\begin{tabular}{|c|c|}
      \hline
      \hbox{\epsfxsize = 6.9 cm  \epsffile{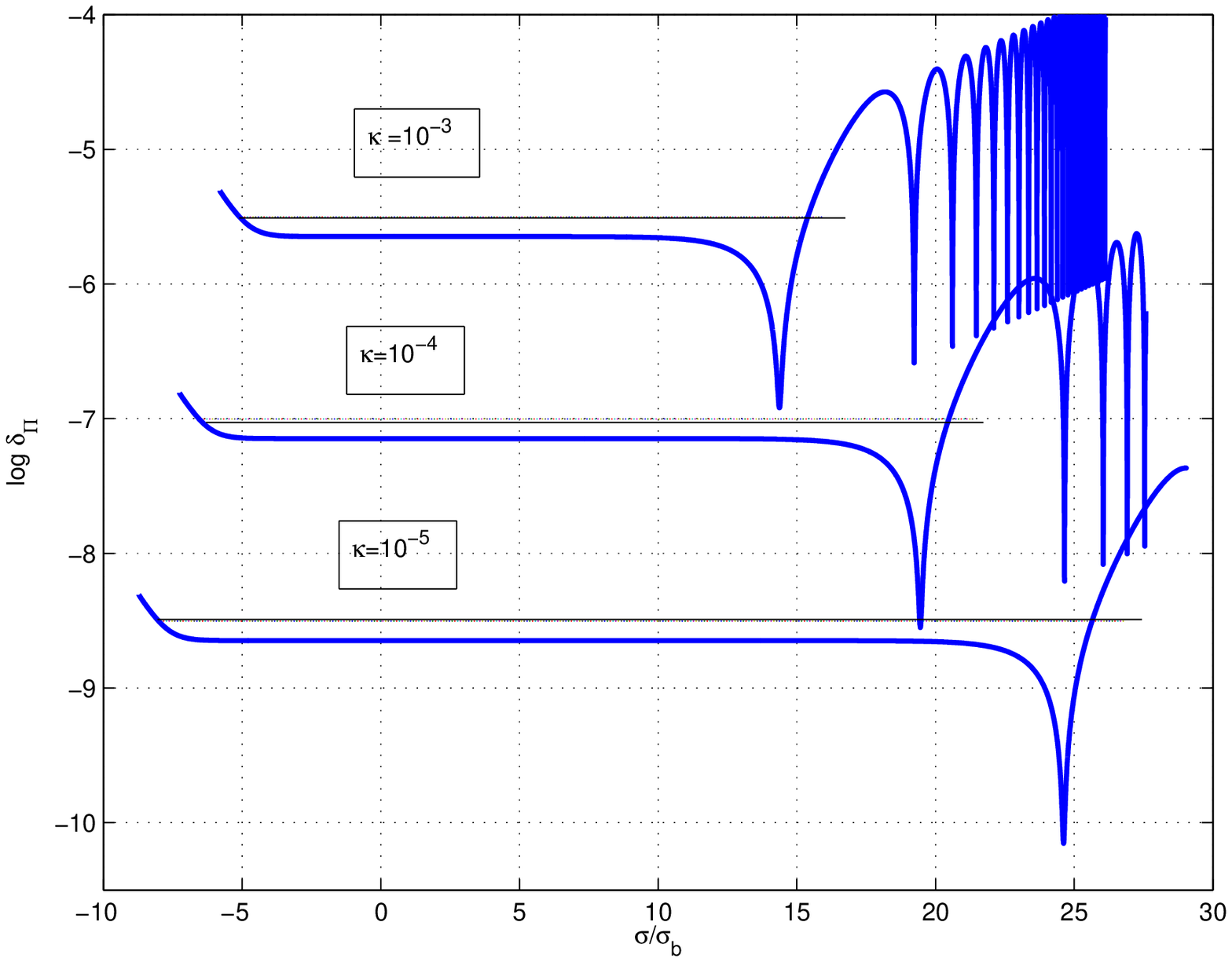}} &
      \hbox{\epsfxsize = 7 cm  \epsffile{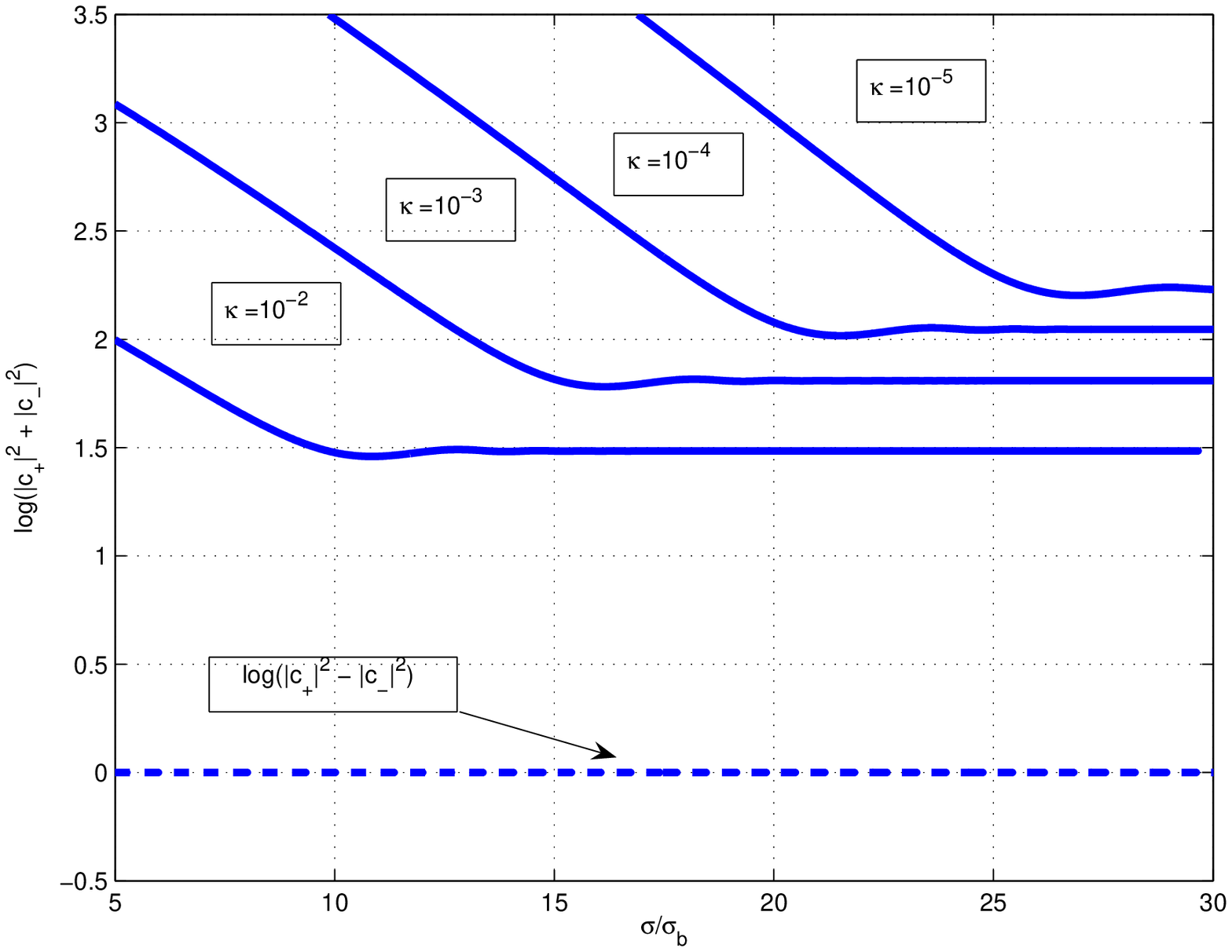}}\\
      \hline
\end{tabular}
\end{center}
\caption{In the plot at the left, the numerical result of the spectrum of the canonical momenta (thick lines), 
$\delta_{\Pi}$ is compared with the analytical expectation (thin lines) for different 
values of the wave-numbers.
In the right plot, the numerical evaluation of the mixing coefficients reported for 
$\kappa \ll 1$ ( see Eqs. (\ref{DIFF}) and (\ref{SUM})).
In both plots the logarithm (to base $10$) of the corresponding function 
is  illustrated as a function 
of $\sigma$ in units of bouncing time $\sigma_{\rm b}$.  The background used 
for this integration corresponds to the case $\alpha=4$, $\phi_{\rm b}=0$ and $\sigma_{1}=\sigma_{2}=0$.} 
\label{figure3}
\end{figure}

Consequently, from Eq. (\ref{MFIII}) and from the second relation reported in Eq. (\ref{AB}) we get 
\begin{equation}
\delta_{\Pi}(k,\sigma) = \frac{k^{3/2}}{2\pi}\sqrt{\frac{{\cal F}_{\rm ex}^2 + \Omega_{\rm ex}^2}{\Omega_{\rm ex}}},
\label{SPPI}
\end{equation}
where the quantities at the right hand side are fully determined analytically
by using Eqs. (\ref{exit}) and (\ref{AB}) and by recalling 
Eq. (\ref{Fsol}).
In Fig. \ref{figure3} (left panel), the analytical expectation given 
by Eq. (\ref{SPPI}) is compared with the numerical result obtained by 
integrating Eqs. (\ref{modefun}) with initial conditions 
dictated by Eq. (\ref{MFII}).
In Fig. \ref{figure3} (plot at the right) the numerical result for the integration 
of Eqs. (\ref{modefun}) with initial conditions 
dictated by Eq. (\ref{MFII}) is reported in terms of the mixing 
coefficients. We checked that, within the accuracy of the 
algorithm, the system of Eqs. (\ref{ss1}), (\ref{ss2}) and (\ref{ss3}) 
with initial conditions given by Eq. (\ref{ssinc}) reproduces 
the same asymptotic results for the mixing coefficients. 
From the analytical estimate the spectrum 
of the mixing coefficients, in the limit $\sigma \gg \sigma_{\rm b}$, leads to 
\begin{equation}
|c_{+}(k)|^2 + |c_{-}(k)|^2 \simeq \frac{11}{4} + \frac{25}{16} \ln^2{\kappa}
\label{cpmanal}
\end{equation}
that agrees with the explicit numerical calculation of the same 
quantity reported in Fig. \ref{figure3}. This statement can be explicitly 
verified by inserting different values of $\kappa$ and by comparing the obtained 
results with numerical values illustrated in the right plot of Fig. \ref{figure3}.
For instance,
for $\kappa =10^{-2}$ we get, from Eq. (\ref{cpmanal}), $\log{[ |c_{+}(k)|^2 + |c_{-}(k)|^2 ]}\simeq 1.55$, for $\kappa \sim 10^{-5}$ we get $\log{[|c_{+}(k)|^2 + |c_{-}(k)|^2 ]}\simeq 2.13$ and so on.

\begin{figure}
\begin{center}
\begin{tabular}{|c|c|}
      \hline
      \hbox{\epsfxsize = 6.9 cm  \epsffile{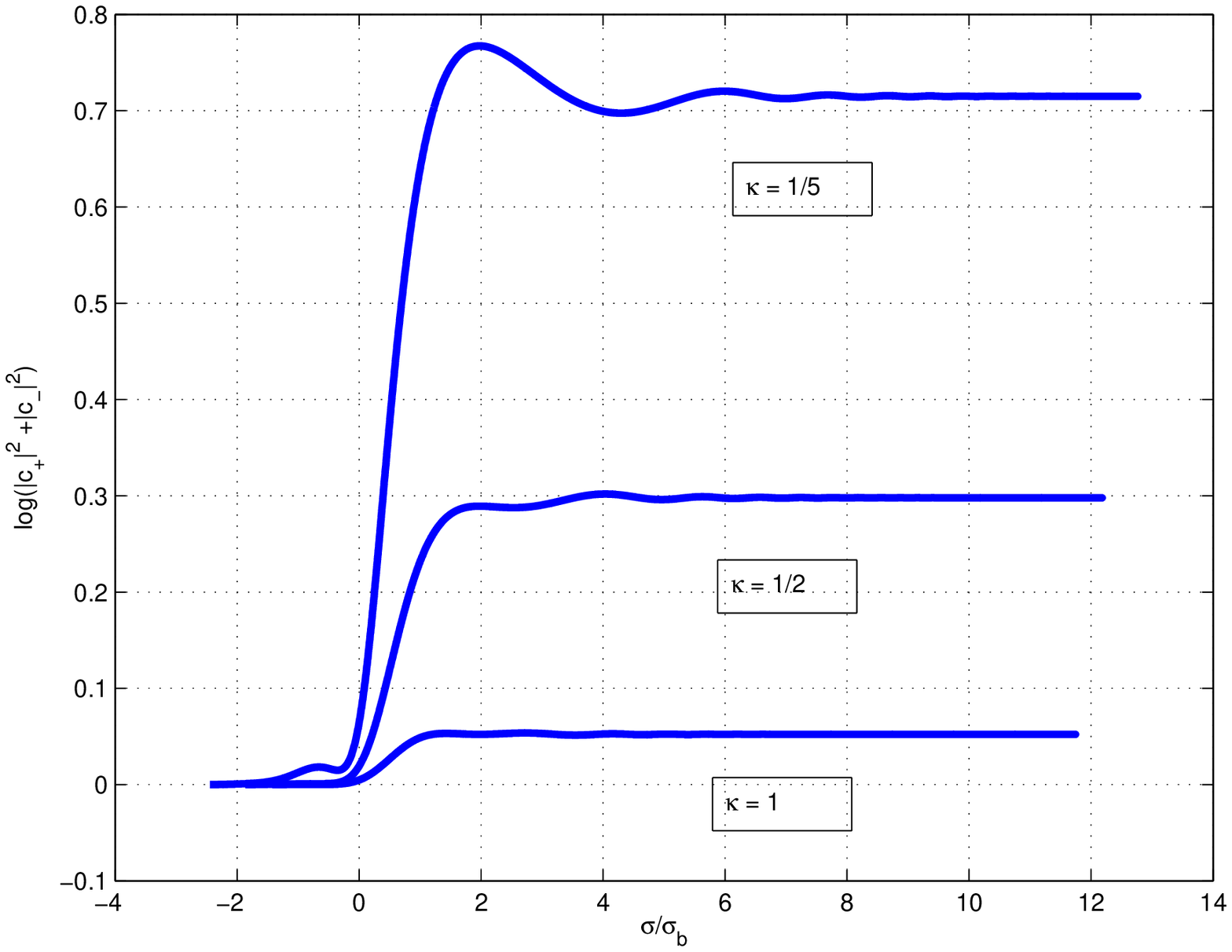}} &
      \hbox{\epsfxsize = 7 cm  \epsffile{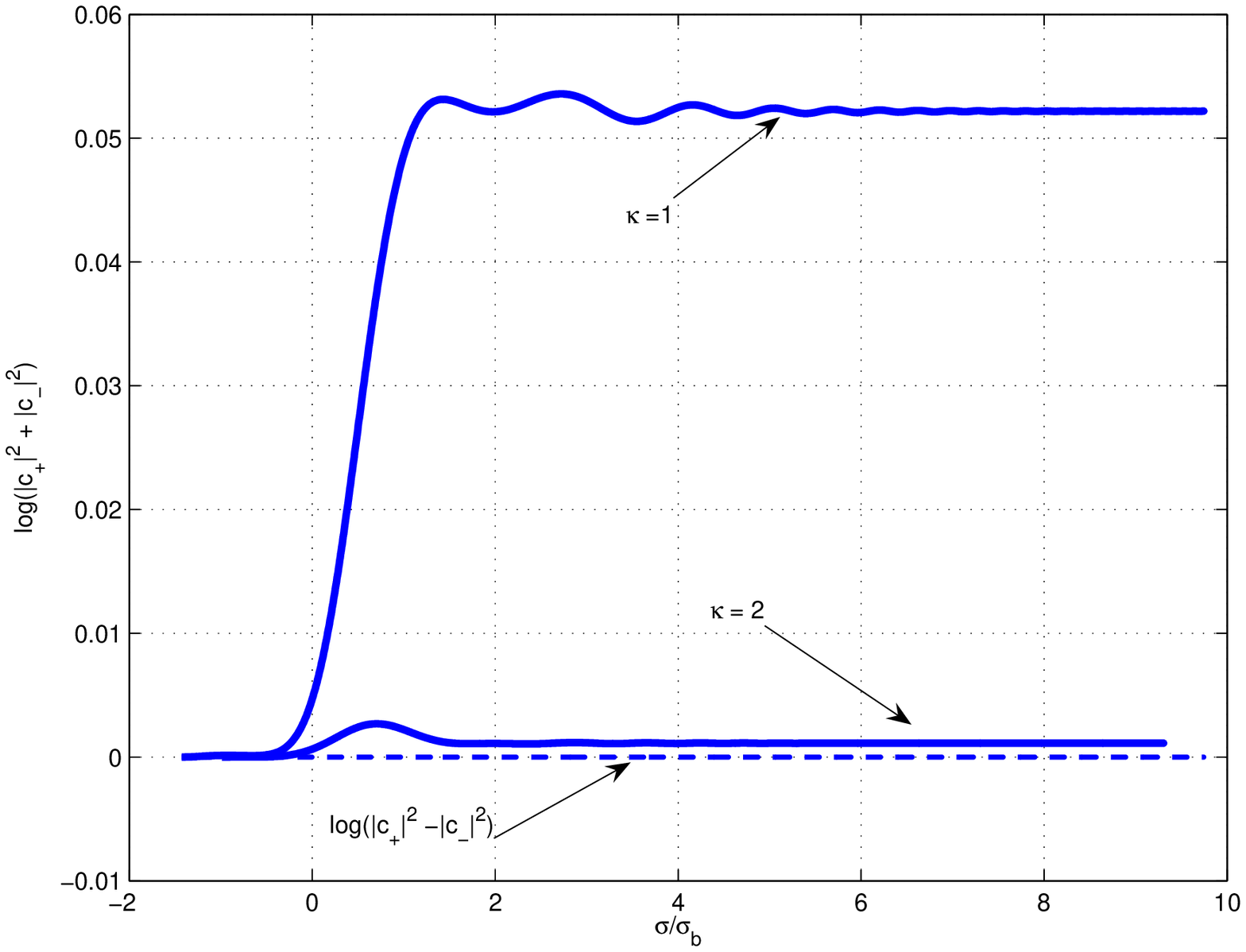}}\\
      \hline
\end{tabular}
\end{center}
\caption{The suppression of the mixing coefficients is illustrated for the 
regime $\kappa > 1$. It is clear from both plots that when $\kappa$ increases 
from $\kappa \leq 1$ to $\kappa >1$, $|c_{-}(k)|^2 \to 0$. The background 
parameters are the same as in Fig. \ref{figure3}.}
\label{figure4}
\end{figure}

While the right plot of Fig. \ref{figure3} nicely illustrates the case 
$\kappa \ll 1$, Fig. \ref{figure4} deals with the case $\kappa \simeq 1$. 
In this regime, the mixing coefficient $c_{-}(k)$ is known 
to be exponentially suppressed. 
It is clear that from the plot at the right that 
as $\kappa$ approaches $1$ the mixing is suppressed in such 
a way that for $\kappa \sim 1$ we have that $(|c_{+}(k)|^2 + 
|c_{-}(k)|^2) \simeq (|c_{+}(k)|^2 - |c_{-}(k)|^2)$ implying 
that $|c_{-}(k)|^2 \simeq 0$.

A relevant implication of the results reported so far is that since the spectrum 
of the mixing coefficients is logarithmically increasing with the wave-number
(for $\kappa = k \sigma_{\rm b} < 1$), 
the spectra of the field operators i.e. $\delta_{h} \simeq k^{3/2} |F_{k}(\sigma)|$ (related to the Fourier transform of the 
two-point function at equal times) will be violet (up to logarithmic 
corrections). By parametrizing the spectrum as 
$\delta_{h} \simeq k^{(n_{\rm t} -1)/2}$ 
we have, in the example discussed so far, $n_{\rm t} =4$ 
(up to logarithmic corrections).
Moreover, since the spectrum of the mixing coefficients increases with frequency it is plausible to expect that the energy spectrum will be even steeper as a function of the wave-number  $k$. 
Consequently, the energy spectrum will be convergent in the 
infrared. In the ultraviolet, the maximally amplified $\kappa$, i.e. 
$\kappa \sim 1$ will give a natural ultraviolet cut-off of the energy density 
when the (initial) zero-point energy is appropriately subtracted. 

\subsection{Energy and pressure densities of relic gravitons}

Different possibilities for assigning the energy and pressure 
densities of the relic gravitons will now be examined.
The first strategy is to compute the
 second-order corrections to the Einstein tensor ${\cal G}_{\mu}^{\nu}$ 
 that appears in Eq. (\ref{EQ1}). The details of the calculation 
 are reported in the appendix and the main result for the components 
 of the energy-momentum pseudo-tensor can be expressed, in the conformal 
 time parametrization, as follows:
\begin{eqnarray}
&&{\cal T}_{0}^{0} =  \frac{1}{a^2 \ell_{\rm P}^2} \biggl[ {\cal H} 
h_{k\ell }' h^{k\ell} + \frac{1}{8} ( \partial_{m} h_{k\ell} \partial^{m} h^{k\ell} + 
h_{k\ell}' {h^{k\ell}}')\biggr],
\label{PSsec1}\\
&& {\cal T}_{i}^{j} = \frac{{\cal T}}{3} \delta_{i}^{j} + \Sigma_{i}^{j}, 
\label{PSsec2}
\end{eqnarray}
where 
\begin{eqnarray}
&& {\cal T} = \frac{1}{a^2 \ell_{\rm P}^2}\biggl[ \frac{5}{8} h_{k\ell}' {h^{k\ell}}' - \frac{7}{8} 
\partial_{m} h_{k\ell} \partial^{m} h^{k\ell} \biggr],
\label{TRsec}\\
&& \Sigma_{i}^{j} = 
\frac{1}{a^2 \ell_{\rm P}^2} \biggl\{ \frac{1}{6} \biggl[ h_{k\ell}' {h^{k\ell}}' - 
\frac{1}{2} \partial_{m} h_{k\ell} \partial^{m} h^{k\ell} \biggr] \delta_{i}^{j}
+ \frac{1}{2} \partial_{m} h_{\ell i} \partial^{m} h^{\ell j} - 
\frac{1}{4} \partial_{i} h_{k\ell} \partial^{j}  h^{k\ell}
\nonumber\\
&& - 
\frac{1}{2} h_{k i}' {h^{k j}}' \biggr\},
\label{ANISsec}
\end{eqnarray}
 with $\Sigma_{i}^{i} =0$.   From Eqs. (\ref{PSsec1}) and (\ref{TRsec})
 the components of the energy and pressure density can be easily 
 obtained since, by definition, $\rho_{\rm gw}= {\cal T}_{0}^{0}$ and 
 $p_{\rm gw} = - {\cal T}/3$. 
 
The components of the energy-momentum 
pseudo-tensor given in Eqs. (\ref{PSsec1}) and (\ref{PSsec2}) 
are not covariantly conserved. However,  since the 
Bianchi identity $\nabla_{\mu} {\cal G}_{\nu}^{\mu}=0$
 should be valid to all orders, we will also have that:
\begin{equation}
\delta_{\rm t}^{(2)} ( \nabla_{\mu} {\cal G}^{\mu}_{\nu}) =0,
\label{bianchi1}
\end{equation}
whose explicit form is 
\begin{eqnarray}
&&\partial_{\mu} \delta_{\rm t}^{(2)} {\cal G}^{\mu}_{\nu} + 
\delta_{\rm t}^{(2)} \Gamma_{\mu\alpha}^{\mu} \overline{\cal G}_{\nu}^{\alpha} + 
\overline{\Gamma}_{\mu\alpha}^{\mu} \delta^{(2)}_{\rm t} {\cal G}_{\nu}^{\alpha} +
\delta_{\rm t}^{(1)} \Gamma_{\mu\alpha}^{\mu} \delta^{(1)}_{\rm t} {\cal G}_{\nu}^{\alpha} 
\nonumber\\
&&- \delta_{\rm t}^{(2)} \Gamma_{\nu\alpha}^{\beta} \overline{{\cal G}_{\beta}^{\alpha}} - 
\overline{\Gamma}_{\nu\alpha}^{\beta} \delta_{\rm t}^{(2)} {\cal G}_{\beta}^{\alpha} -
\delta_{\rm t}^{(1)} \Gamma_{\nu\alpha}^{\beta} \delta^{(1)}_{\rm t} {\cal G}_{\beta}^{\alpha}=0.
\end{eqnarray}
Recalling  now the components of the energy-momentum 
pseudo-tensor  and the results for the fluctuations of the Christoffel symbols 
(i.e. Eqs. (\ref{CONN}) of the appendix) we have 
\begin{equation}
\frac{\partial \rho_{\rm gw}}{\partial\tau} + 3 {\cal H}( \rho_{\rm gw} + p_{\rm gw}) 
- \frac{2 ({\cal H}^2 - {\cal H}')}{a^2 \ell_{\rm P}^2} \delta_{\rm t}^{(2)} \Gamma_{k0}^{k}=0,
\end{equation}
that can also be written as 
\begin{equation} 
\frac{\partial \rho_{\rm gw}}{\partial\tau} + 3 {\cal H}( \rho_{\rm gw} + {\cal P}_{\rm gw}) =0
\label{conteq}
 \end{equation}
 where 
\begin{equation}
 {\cal P}_{\rm gw} =  p_{\rm gw} +  \frac{({\cal H}^2 - {\cal H}')}{ 3 {\cal H} a^2} h_{k\ell}' h^{k\ell}.
 \label{Pmod1}
\end{equation}
Equations (\ref{PSsec1}), (\ref{TRsec}) and (\ref{Pmod1}) imply that the operators corresponding to the energy 
 and pressure densities are, in the $\sigma$-parametrization,
\begin{eqnarray}
 \hat{\rho}_{\rm gw} &=& \frac{1}{a^6}\biggl\{ 
4 {\cal F}[\hat{h} \hat{\Pi} + \hat{\Pi} \hat{h}] + a^4 \partial_{m} \hat{h} \partial^{m} \hat{h} 
+ \hat{\Pi}^2\biggr\},
\label{oprho}\\
\hat{p}_{\rm gw} &=&  \frac{1}{3 a^6} \biggl[ - 5 \hat{\Pi}^2 
+ 7 a^4 (\partial_{m} \hat{h}) (\partial^{m} \hat{h})\biggr],
\label{opp}\\
\hat{{\cal P}}_{\rm gw} &=& \hat{p}_{\rm gw} 
+ \frac{4}{3 a^6 } \biggl({\cal F} - \frac{d \ln{{\cal F}}}{d\sigma}\biggr) 
[\hat{\Pi} \hat{h}  + \hat{h}  \hat{\Pi}],
\label{opP}
\end{eqnarray}
where $\hat{h}$ and $\hat{\Pi}$ are the canonical field operators 
defined in Eq. (\ref{oper}).
The averaged components of the energy-momentum pseudo-tensor 
can be obtained by taking the expectation values of the operators defined in Eqs.  (\ref{oprho}), (\ref{opp}) and (\ref{opP}) and by recalling that the initial state 
is the one annihilated by the creation and destruction operators of Eqs. (\ref{Heis1a}) 
and (\ref{Heis1b}). The result is then expressed in terms of the appropriate 
mode functions, i.e. 
\begin{eqnarray} 
&& \langle \hat{\rho}_{\rm gw} \rangle= \frac{1}{a^{6}} \int \frac{d^{3} k}{(2\pi)^3} \biggl\{ 
4 {\cal F} [ F_{k}(\sigma) G_{k}^{*}(\sigma) + F_{k}^{*}(\sigma) G_{k}(\sigma)] + \Omega^2 |F_{k}(\sigma)|^2 
+ |G_{k}(\sigma)|^2\biggr\},
\label{secrho}\\
&& \langle \hat{p}_{\rm gw} \rangle=  \frac{1}{3 a^6} \int \frac{d^{3} k}{(2\pi)^3} [ 7 \Omega^2 |F_{k}(\sigma)|^2 - 5 |G_{k}(\sigma)|^2],
\label{secp}\\
&& \langle \hat{{\cal P}}_{\rm gw} \rangle = \langle \hat{p}_{\rm gw} \rangle+ 
\frac{4}{3 a^{6}} \int \frac{d^{3} k}{(2\pi)^3}\biggl( 3 {\cal F} - 
\frac{d \ln{{\cal F}}}{d\sigma}\biggr) [ F_{k}(\sigma) G_{k}^{*}(\sigma) + F_{k}^{*}(\sigma) G_{k}(\sigma)].
\label{secP}
\end{eqnarray}
It is relevant to remark here that Eq. (\ref{modefun}) implies that $ \partial_{\sigma} \langle \hat{\rho}_{\rm gw} \rangle + 3 {\cal F}  \langle \hat{\rho}_{\rm gw} + \hat{{\cal P}}_{\rm gw}\rangle =0$.
The initial conditions for the field operators  (\ref{Heis1a}) and (\ref{Heis1b}) for $\sigma_{\rm i}\to 
-\infty$ imply that the contribution of the zero-point energy is 
given by $\int (k^4)/(2\pi^2 a^4) d\ln{k}$ (where we recalled that $\Omega
= k\,a^2$). To discard this quantity by appropriate 
subtraction (both in the energy and pressure densities) amounts 
to neglect the contribution of the zero-point oscillations 
of the vacuum.
As already mentioned, there are different 
ansatz for the energy-momentum pseudo-tensor that 
have been proposed in order to treat back-reaction effects 
of the relic gravitons. It is therefore appropriate 
to comment here about these possiblities.
By looking at the form of Eq. (\ref{TA1}), the authors of Ref. \cite{ford1} (see 
also \cite{ford2}) argued that a natural ansatz for the energy and pressure 
densities of the relic gravitons is the one we can derive from the energy-momentum 
tensor of a minimally coupled scalar field for each of the two tensor 
polarizations.
This consideration implies that the energy and pressure 
densities can be written as 
\begin{eqnarray}
&& \hat{\rho}_{\rm gw} = \frac{1}{a^6} [ \hat{\Pi}^2 
+ a^4 \partial_{m} \hat{h} \partial^{m} \hat{h}],
 \label{rh2}\\
 && \hat{p}_{\rm gw} =  \frac{1}{a^6} \biggl[ \hat{\Pi}^2  -
 \frac{a^4}{3} \partial_{m} \hat{h} \partial^{m} \hat{h} \biggr].
 \label{p2}
 \end{eqnarray}
The averaged energy and pressure densities become then:
\begin{eqnarray}
&& \langle \hat{\rho}_{\rm gw} \rangle = \frac{1}{a^6} \int \frac{d^{3} k}{(2\pi)^3} 
[ |G_{k}(\sigma)|^2 + 
\Omega^2 |F_{k}(\sigma)|^2 ],
\label{FPrho}\\
&&\langle \hat{p}_{\rm gw} \rangle= \frac{1}{a^6} \int \frac{d^{3} k}{(2\pi)^3}\biggl[ |G_{k}(\sigma)|^2 - \frac{\Omega^2}{3} |F_{k}(\sigma)|^2 \biggr].
\label{FPp}
\end{eqnarray}
Equations (\ref{FPrho}) and (\ref{FPp})  imply $ \partial_{\sigma} \langle \hat{\rho}_{\rm gw} \rangle + 3 {\cal F}  \langle \hat{\rho}_{\rm gw} + \hat{p}_{\rm gw}\rangle =0$
provided the evolution equations of $F_{k}$ and $G_{k}$ are the ones  given in Eq. (\ref{modefun}). 

In order to keep track of the possible ambiguities 
related with the definitions of the energy-momentum 
pseudo-tensor we will first perform the calculations 
within the approach defined by  Eqs. (\ref{secrho}), (\ref{secp}) and (\ref{secP}). Then, 
for each dynamical quantity,  the results will be compared with the answers 
obtained using the approach defined by Eqs. (\ref{FPrho}) and (\ref{FPp}).

\renewcommand{\theequation}{4.\arabic{equation}}
\section{Effective barotropic indices}
\setcounter{equation}{0}
As a consequence of the dynamical evolution of the background 
the energy and pressure densities have a steep (violet) spectrum. 
Using the results of the previous section, it appears, for instance 
that  for $ k < k_{\rm max}$ we will have, qualitatively,
that the logarithmic spectrum of the energy density is \footnote{In the 
following sections, for sake of simplicity, the averages $\langle \hat{\rho}_{\rm gw}
\rangle$ will be denoted simply by $\rho_{\rm gw}$ and similarly for the 
the pressure densities.}
\begin{equation}
a^4 \frac{d \rho_{\rm gw}}{d\ln{k}} \simeq k^4 |c_{-}(k)|^2,
\end{equation}
where the initial zero-point energy has been already subtracted. 
For $k > k_{\rm max}$ the mixing coefficient 
is not enhanced logarithmically, but rather exponentially suppressed (see 
Fig. \ref{figure4} where this aspect is numerically illustrated).
Consequently it is 
plausible that the integrals appearing in the averaged energies 
and pressure densities are dominated by the modes that 
are maximally amplified, i.e. $\kappa \simeq \kappa_{\rm max}$.
Since $\kappa_{\rm max}\simeq 1$, $k_{\rm max} = 1/\sigma_{\rm b}$ (see 
Fig. \ref{figure4}). The direct calculation supports this view since 
 integrals of the energy density can be performed 
numerically. Let us then define, for practical reasons, 
\begin{equation}
a^{6} \rho_{\rm gw} = \int^{k_{\rm max}} {\cal A}_{1,2}(k,\sigma) d\ln{k},
\label{defA12}
\end{equation}
where the indices $1$ and $2$ refer to the two different 
parametrizations of the energy density of the relic gravitons discussed 
respectively, in Eqs. (\ref{secrho}) and (\ref{FPrho}).
\begin{figure}
\begin{center}
\begin{tabular}{|c|c|}
      \hline
      \hbox{\epsfxsize = 6.9 cm  \epsffile{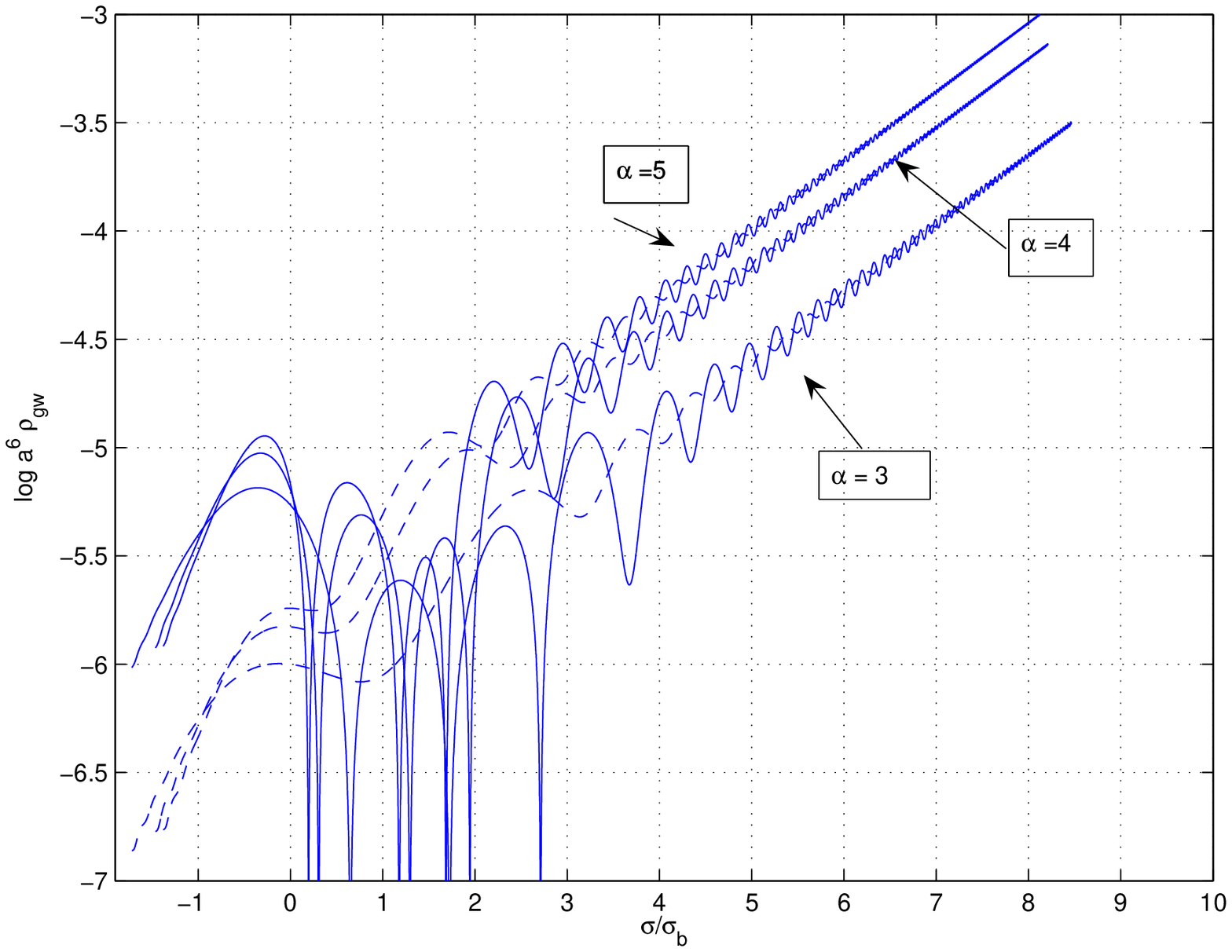}} &
      \hbox{\epsfxsize = 7 cm  \epsffile{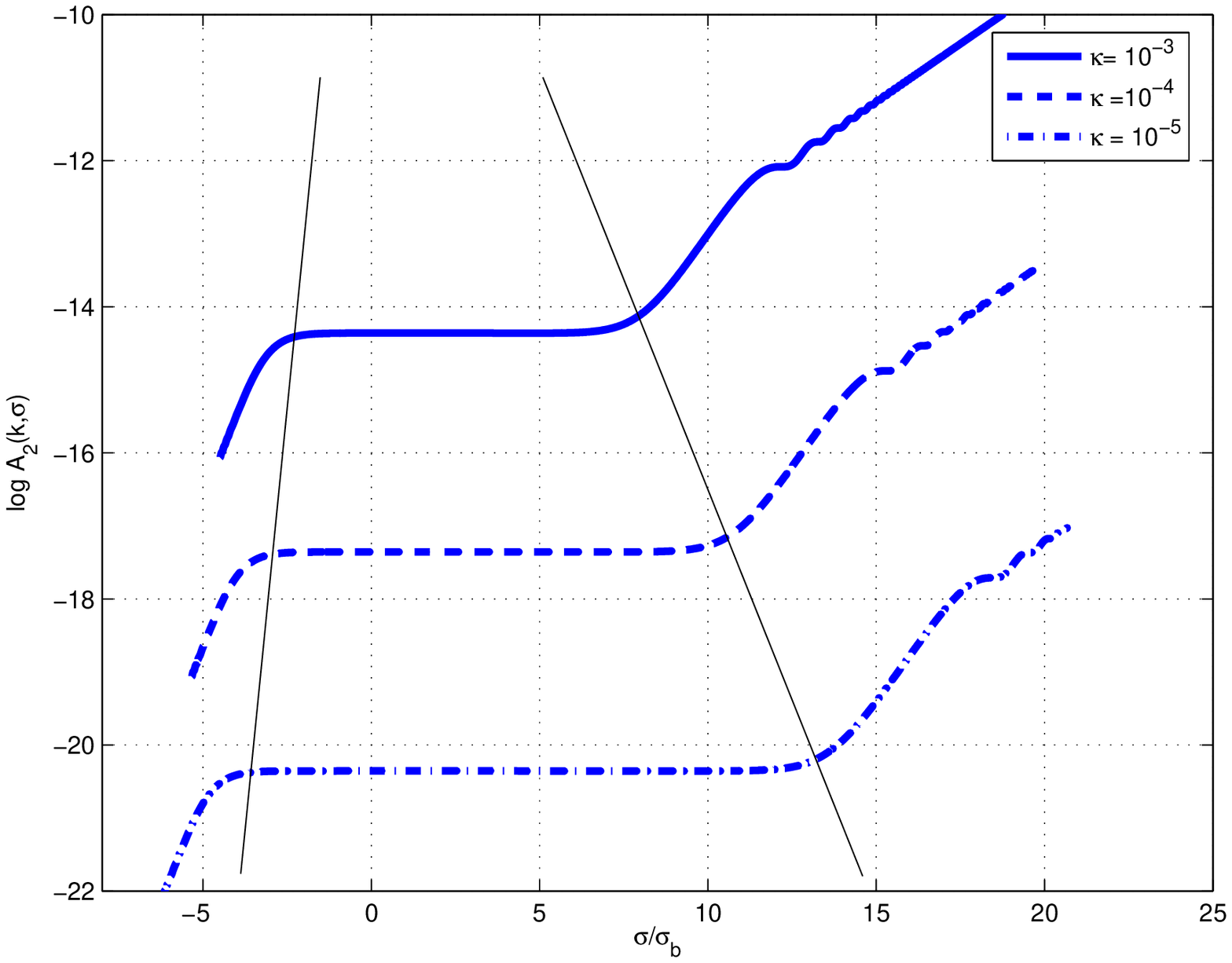}}\\
      \hline
\end{tabular}
\end{center}
\caption{Numerical evaluation (left plot) of the integrated energy density 
for different values of $\alpha$. 
The contribution of a sample of $k$-modes for the case 
$\alpha =4$. Initial conditions for the background are fixed in such a 
way that $\sigma_{1}=\sigma_2=0$, while the gauge coupling at the bounce 
is of order $1$. The initial conditions for the tensor modes are the ones 
dictated by quantum mechanics (see Section 3).}
\label{figure5}
\end{figure}
In Fig. \ref{figure5} (left plot) the numerical evaluation 
of $\rho_{\rm gw}$ is reported. The initial conditions for 
the mode functions are the ones specified in the previous section 
for eack $k$-mode. With the dashed line the energy density 
discussed in Eq. (\ref{FPrho}) is illustrated. With the full line 
the energy density of Eq. (\ref{secrho}) is instead reported.
The integral appearing in Eq. (\ref{defA12}) is indeed dominated by the upper 
limit. This aspect is also illustrated, in a related perspective 
from the right plot of Fig. \ref{figure5} where the contribution 
of the integrand for $\kappa \ll 1$ is reported for a sample of $k$-modes.
In practice, it is useful to specify a given $\kappa_{\rm min}$, i.e. the lower 
limit of integration appearing in Eq. (\ref{defA12}).  For instance, typical values 
chosen for the present examples are $\kappa = 10^{-12}$ or $\kappa =10^{-10}$. Since the spectrum is  dominated by $\kappa_{\rm max}$, different choices $\kappa_{\rm min}$ produce the same results for the integrated energy density. According to Eq. (\ref{exit}), $\kappa_{\rm min}$ determines 
a minimal $\sigma_{\rm ex}$, i.e. $\sigma_{\rm ex}(\kappa_{\rm min})$. This 
implies that, to be consistent, the initial conditions for the field operators have been given for $\sigma_{\rm i}\ll \sigma_{\rm ex}(\kappa_{\rm min})$ 
so that, at the initial time, all the tensor modes were of short wavelength.

The right panel of Fig. \ref{figure5} has some interesting features that will now 
be discussed. Within the region limited by the two diagonal lines the corresponding modes satisfy $|\kappa \tau| <1$. Bearing in mind 
Eq. (\ref{defA12}), we then have that for modes of long wavelength
 $a^6 \rho_{\rm gw} $ is constant.  
 Outside the region marked by the two diagonal lines the corresponding 
 modes satisfy $|\kappa \tau| >1$. In this region $a^6 \,\rho_{\rm gw}$ 
 increases as $a^2$, i.e.  $\rho_{\rm gw} \sim a^{-4}$. 
 
These considerations suggest that in the short wavelength limit the effective barotropic index 
is the one of radiation, i.e.  $1/3$. In fact, in the short
wavelength limit the approximate evolution of the mode functions 
is simply given by Eqs. (\ref{MFI}) and (\ref{MFI}). Inserting these 
solutions either in Eqs. (\ref{secrho})--(\ref{secP}) or in Eqs. (\ref{FPrho})--(\ref{FPp})
the results are the same for both parametrizations. 
In the long wavelength limit, according to Fig. \ref{figure5} we should instead 
have that the effective barotropic index is close to $1$.
 
To corroborate the numerical result, we can solve, directly in the conformal 
time parametrization, the evolution 
equations of the mode functions in the approximate background 
of Eqs. (\ref{anal1}) and (\ref{phanal}). The (averaged) energies 
and pressure densities will then be computed and the  result  expanded 
in the limit of $k\tau \ll 1$.

Following this procedure (see appendix B and, in particular Eqs. (\ref{prev}) and (\ref{prev2})), from Eqs. (\ref{FPrho}) and (\ref{FPp}),
we have for $\tau <- \tau_{1}$:
\begin{eqnarray}
&& \rho_{\rm gw}  \simeq  \frac{1}{a^4}\int \frac{k^4}{2\pi^2} 
d\ln{k} \biggl[ \frac{1}{(-\pi k\tau)} + {\cal O}(k\tau)\biggr],\qquad \tau <-\tau_{1}
\nonumber\\
&& p_{\rm gw}  \simeq  \frac{1}{a^4}\int \frac{k^4}{2\pi^2} 
d\ln{k} \biggl[ \frac{1}{(-\pi k\tau)} + {\cal O}(k\tau)\biggr],\qquad \tau< -\tau_{1}.
\label{lwFP}
\end{eqnarray}
Thus, from Eqs. (\ref{lwFP}) we have that,  for $k \tau \ll 1$, $p_{\rm gw} = \rho_{\rm gw}$.
 Furthermore, recalling that, for $\tau < -\tau_{1}$, $a(\tau) \sim \sqrt{-\tau}$, we have
 as expected that $\rho_{\rm gw}$ and $p_{\rm gw} $ scale as $a^{-6}$.
 
If the energy-momentum pseudo-tensor is derived from the quadratic 
corrections to the Einstein tensor, the ratio between the pressure and the 
energy densities becomes, for $k\tau \ll 1$, 
\begin{equation}
\frac{ {\cal P}_{\rm gw} }{  \rho_{\rm gw} } 
= 1 - \frac{5}{3} \frac{1}{ [1 - 4 \gamma + 4 \ln{2} - 4 \ln{(- k\tau)}]},
\label{lwsec}
\end{equation}
where $\gamma$ is the Euler-Mascheroni constant. Notice 
that this result is consistent with the one of Eqs. (\ref{lwFP}) with the difference 
that in Eq. (\ref{lwsec}) logarithmic corrections do appear.

It is instructive to compare the present case with the one of a sudden transition
from a de Sitter stage of expansion to a radiation-dominated stage of expansion. The typical time-scale $\tau_{1}$ will now mark the transition from 
a de Sitter stage of expansion to a radiation-dominated epoch and the 
analytical form of the scale factor in the two regions is:
\begin{eqnarray}
&& a(\tau) = - \frac{\tau_{1}}{\tau},\qquad \tau \leq - \tau_{1},
\label{dsa}\\
&& a(\tau) = \frac{\tau + 2 \tau_{1}}{\tau_{1}}, \qquad \tau > -\tau_1.
\label{rada}
\end{eqnarray}
In this case (see appendix B for further details), Eqs. (\ref{secrho})-(\ref{secP}) imply, for 
$\tau < -\tau_{1}$,
\begin{eqnarray}
&&  \rho_{\rm gw}  = \frac{1}{a^4} \int \frac{k^4}{2\pi^2} \,\, 
d\ln{k}\,\biggl[
1 - \frac{7}{2 x^2}\biggr],
\label{exdsa1}\\
&& {\cal P}_{\rm gw}= p_{\rm gw}  = \frac{1}{3 a^4} \int \frac{k^4}{2\pi^2}\,\, 
d\ln{k} \biggl[
1 + \frac{7}{2 x^2}\biggr],
\label{exdsa2}
\end{eqnarray}
where $ x = k\tau$.
The energy-density is positive for $k\tau \gg 1$ but it becomes 
negative when the relevant modes become larger than the Hubble radius, 
i.e. for $k\tau \ll 1$. The equation of state has an effective barotropic index $1/3$,  for  $k\tau \gg 1$. In the opposite limit, i.e. $k \tau \ll 1$ the effective barotropic index 
becomes$ -1/3$. 
From Eqs. (\ref{exdsa1}) and (\ref{exdsa2}) the energy and pressure 
densities scale as $a^{-2}$  and this 
is consistent with Eq. (\ref{conteq}) for the equation of state 
$ {\cal P}_{\rm gw}  = - \rho_{\rm gw}  / 3$.

Consider then the case of modes $k \tau \ll 1$ for $\tau > -\tau_{1}$ 
i.e.  during the radiation-dominated epoch.  The exact results are reported 
in the appendix and here only the long wavelength limit will be given. 
As as the approach defined by Eqs. (\ref{TRsec})--(\ref{ANISsec}) is concerned, the relevant results are reported in Eqs. (\ref{AVrho1}), (\ref{AVp1}) and (\ref{AVP1}) of the appendix. To lowest order in $|k\tau|\ll 1$ and 
$|k\tau_1|\ll 1$ Eq. (\ref{AVrho1}) leads to 
\begin{equation}
 \rho_{\rm gw}  = - \frac{5}{6} \frac{1}{a^2} 
\int \frac{d\ln{k}}{2\pi^2 \tau_{1}^4} k^2.
\label{eff1}
\end{equation}
In the same limit, from Eqs. (\ref{AVp1}) the following result
\begin{equation}
p_{\rm gw}  = \frac{7}{6} \frac{1}{a^2} 
\int \frac{d\ln{k}}{2\pi^2 \tau_{1}^4} k^2,
\label{eff2}
\end{equation}
can be obtained.
We can immediately notice that $ p_{\rm gw}/ \rho_{\rm gw} = - \frac{7}{5}$,
which agrees with the calculation of \cite{abramo1,abramo2}. The fact that 
the averaged energy density can have negative values for a limited 
amount of time was noticed, in a related context, in \cite{kuo}. 

Finally, from Eq. (\ref{AVP1}),
to lowest order in $|x | \ll 1$ and $|x_1|\ll 1$,
\begin{equation}
{\cal P}_{\rm gw} - p_{\rm gw}  = -\frac{8}{9} \frac{1}{a^2} 
\int \frac{d\ln{k}}{2\pi^2 \tau_{1}^4} k^2.
\label{eff2a}
\end{equation}
Therefore, putting together the results expressed in Eqs. (\ref{eff1}), (\ref{eff2})
and (\ref{eff2a}), the following chain of equalities holds
\begin{eqnarray}
&& p_{\rm gw} = - \frac{7}{5} \rho_{\rm gw} ,
\label{chain0}\\
&&  {\cal P}_{\rm gw}  = - 
\frac{7}{5} \rho_{\rm gw} + 
\frac{16}{15} \rho_{\rm gw}   = - \frac{1}{3} \rho_{\rm gw}. 
\label{chain1}
\end{eqnarray}
These results coincide with the ones obtained in \cite{abramo1,abramo2}. 

Instead of using the energy-momentum pseudo-tensor 
derived from the quadratic corrections to the Einstein tensor, we could use the 
energy and pressure densities discussed in Eqs. (\ref{rh2}) and (\ref{p2}).
The relevant results, in this case, are reported in Eqs. (\ref{AVrho2}) and 
(\ref{AVp2}) of the appendix. Again, to lowest order in $|x | \ll 1$ and $|x_1|\ll 1$, Eqs. (\ref{AVrho2}) and (\ref{AVp2}) lead to 
\begin{eqnarray}
&&  \rho_{\rm gw} = \frac{1}{ 2 a^4} \int \frac{k^4}{2\pi^2} d\ln{k} 
\frac{x^2}{  x_1^4}\biggl[ 1 - \frac{2}{9} x^2 +\,\,...\,\,\biggl],
\label{eff3}\\
&&  p_{\rm gw}  = - \frac{1}{ 6 a^4} \int \frac{k^4}{2\pi^2} d\ln{k} 
\frac{ x^2 }{  x_1^4}\biggl[ 1 - \frac{6}{9} x^2 + \,\,...\,\, \biggl],
\label{eff4}
\end{eqnarray}
where the ellipses stand for higher-order corrections. While in the case 
of Eq. (\ref{eff1}) the averaged energy density is negative, in the case of Eq. 
(\ref{eff3}) the averaged energy density is positive. Moreover, from 
Eqs. (\ref{eff3}) and (\ref{eff4}) we have that the effective equation 
of state in the long wavelength limit is given by 
\begin{equation}
 p_{\rm gw} = - \frac{1}{3}  \rho_{\rm gw},
\label{chain2}
\end{equation}
which is the same of Eq. (\ref{chain1}).

Therefore, we can conclude that both in the bouncing case and in 
the transition from de Sitter to radiation the two parametrizations 
of the energy-momentum pseudo-tensor give the same information. 
In particular, ${\cal P}_{\rm gw}$ defined by means of the corrections 
to the Einstein tensor obeys the same equation of state obeyed, in the 
approach of Eqs. (\ref{rh2}) and (\ref{p2}) by $p_{\rm gw}$. 

\renewcommand{\theequation}{5.\arabic{equation}}
\section{Iterative calculations of back-reaction effects}
\setcounter{equation}{0}
The evolution equations including the dynamical effects 
of the produced gravitons reduce to an integro-differential 
system whose numerical solution may be obtained by means of iterative 
methods that we are now going to describe and exploit.  
One of the important qualitative features of the bounce solutions 
discussed in Section 2 was expressed by the Eq. (\ref{int1}) whose 
form becomes now 
\begin{equation}
\frac{d}{d\sigma} \biggl( {\cal F} + \frac{\xi}{2} \biggr) =
\frac{a^6}{4} ( \rho_{\rm gw} - p_{\rm gw}),
\label{SC1}
\end{equation}
where with $\rho_{\rm gw}$ and $p_{\rm gw}$ we denote the averages of the components 
of the energy-momentum pseudo-tensor of the relic gravitons.  
To this equation the evolution equation 
for $\xi$ should be added, i.e. 
\begin{equation}
\frac{d \xi}{d\sigma} = - e^{\phi} a^6 \biggl[ V - \frac{\partial V}{\partial \overline{\phi}}\biggr],
\label{SC2}
\end{equation}
which is (formally) independent on the energy and pressure 
densities of the created gravitons. 
The dynamical quantities appearing in Eq. (\ref{SC1}) and (\ref{SC2}) 
are subjected to the following (generalized) Hamiltonian 
constraint:
\begin{equation}
{\cal F}^2 = \frac{a^6}{6} \rho_{\rm gw} 
+ \frac{\xi^2}{12} + \frac{e^{\phi} a^6}{6} V,
\end{equation}
that has to be satisfied along the different steps of the numerical solution.
Recalling the examples given in Fig. \ref{figure5}, $\rho_{\rm gw}$ and $p_{\rm gw}$ are computed 
by integrating the evolution equations of the mode functions, i.e.
Eqs. (\ref{modefun}).  In this sense, Eq. (\ref{SC1}) is 
actually an integro-differential equation whose explicit form is indeed
\begin{equation}
\frac{d}{d\sigma}\biggl[ {\cal F} + \frac{\xi}{2}\biggr] = 
\frac{1}{4} \int ^{k_{\rm max}} {\cal B}(\Omega,\sigma) \,\,d\ln{k},
\label{IDS}
\end{equation}
where the quantity ${\cal B}(\Omega,\sigma)$ depends on the specific 
form of the energy-momentum pseudo-tensor. In particular, 
subtracting Eq. (\ref{FPp}) from Eq. (\ref{FPrho}) 
\begin{equation}
{\cal B}(\Omega,\sigma) =  \frac{2 k^3}{3 \pi^2 } [\Omega^2(\sigma) |F_{k}(\sigma)|^2 - \Omega(\sigma)].
\label{IDS2}
\end{equation}
Similarly, taking the difference between Eq. (\ref{secrho}) and (\ref{secp}) 
\begin{equation}
{\cal B}(\Omega,\sigma) = \frac{ 2 k^3}{3 \pi^2}\{2 |G_{k}(\sigma)|^2 - 
\Omega^2(\sigma) |F_{k}(\sigma)|^2 + 3 {\cal F}[ F_{k}(\sigma)^{*} G_{k}(\sigma) 
+ F_{k}(\sigma) G_{k}(\sigma)^{*} - \Omega(\sigma)]\}.
\label{IDS3}
\end{equation}
In both equations $F_{k}(\sigma)$ and $G_{k}(\sigma)$ satisfy Eqs. (\ref{modefun}). Note that in Eqs. (\ref{IDS2}) and (\ref{IDS3}) 
the zero-point energy and pressure densities have been subtracted as discussed in Section 4.

Equation (\ref{IDS}) can be solved by iteration \footnote{It should be mentioned that an iterative approach for the solution 
of the back-reaction problems has been also invoked by the authors of Ref. 
\cite{HUP}.}. In fact, from 
Eq. (\ref{IDS}) the following hierarchy of equations can be deduced:
\begin{eqnarray}
&& 
{\cal F}_{0}(\sigma) + \frac{\xi_{0}(\sigma)}{2}   =0,
\label{IT1}\\
&&  {\cal F}_{1}(\sigma) + \frac{\xi_{1}(\sigma)}{2}  = 
\frac{1}{4}\int^{\sigma} d\sigma' \int_{0}^{k_{\rm max}} {\cal B}_{0}(\Omega(\sigma'), \sigma')\, d\ln{k} ,
\label{IT2}\\
&& {\cal F}_{2}(\sigma) + \frac{\xi_{2}(\sigma)}{2} = \frac{1}{4}\int^{\sigma} d\sigma' \int_{0}^{k_{k_{\rm max}}} {\cal B}_{1}(\Omega(\sigma'), \sigma') \,d\ln{k},
\label{IT3}
\end{eqnarray}
and so on.

The subscripts appearing in the various dynamical  quantities of Eqs. (\ref{IT1}), (\ref{IT2}) 
and (\ref{IT3}) denote the order of the iteration. The source term appearing 
in the hierarchy is computed from the previous order. So, to zeroth order, i.e. 
Eq. (\ref{IT1}), ${\cal F}_{0}$ and $\xi_{0}$  are simply the solutions discussed in Section 2.
These solutions will be used to obtain numerically the mode functions that will 
determine ${\cal B}_{0}(\Omega,\sigma)$.  To first order, ${\cal F}_{1}$ and $\xi_{1}$ will be determined from the the zeroth order source, i.e. ${\cal B}_{0}(\Omega,\sigma)$. This will lead to the first-order mode functions 
allowing the quantitative determination of $ {\cal B}_{1}(\Omega,\sigma)$ and so on.

Based on the results of the previous section, the strongest effect is expected 
from modes of high wave-number (i.e. small wavelength).  
The rationale 
for this expectation is that in the long wavelength limit 
 $a^{6}(\rho_{\rm gw} -p_{\rm gw})$, is smaller than the energy density of the dilaton  and also approximately vanishing. In the short wavelength limit $ a^{6}(\rho_{\rm gw} -p_{\rm gw})$
 increases as $a^2$ and it is expected to dominate the background at sufficiently  late times: as time goes by more and more modes will contribute 
 to the ultraviolet branch of the energy spectrum. 
  For the numerical solution of the 
 problem the (stiff) Rosenbrok method \cite{numrec} has been used.  
 \begin{figure}
\begin{center}
\begin{tabular}{|c|c|}
      \hline
      \hbox{\epsfxsize = 6.9 cm  \epsffile{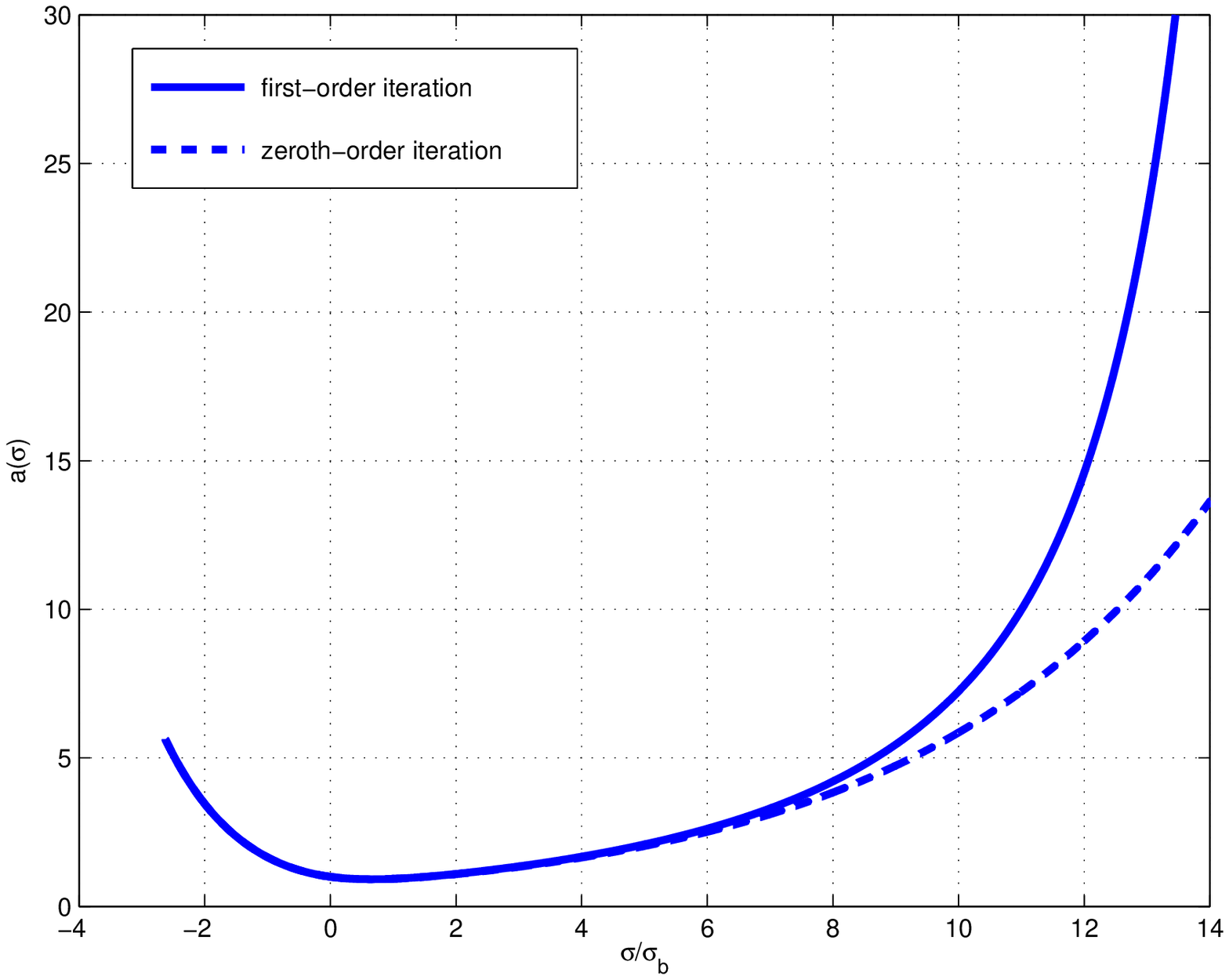}} &
      \hbox{\epsfxsize = 7 cm  \epsffile{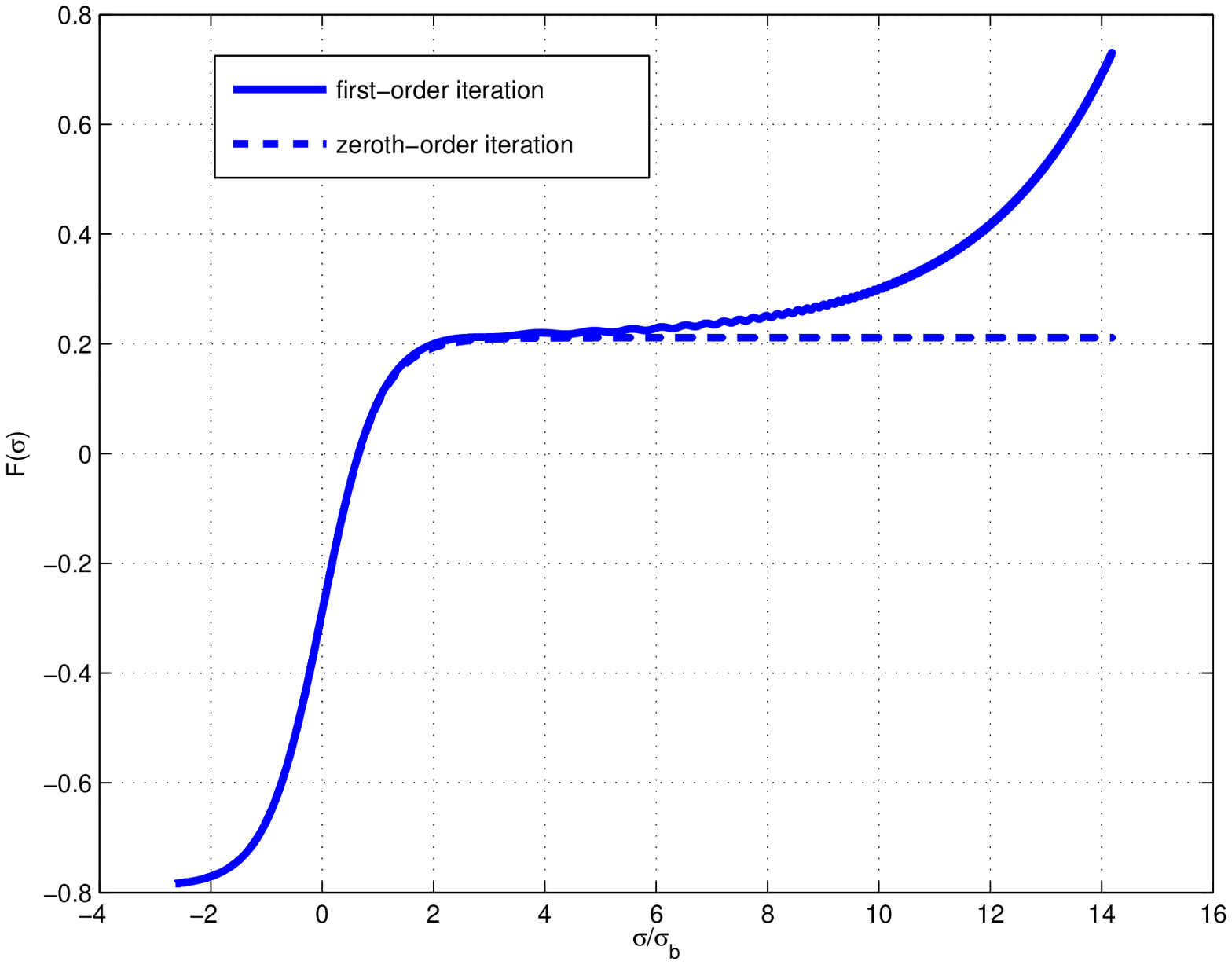}}\\
      \hline
\end{tabular}
\end{center}
\caption{The evolution of the scale factor (left plot) and of ${\cal F}$ (right plot) 
in the case $\alpha =4$. The back-reaction effects are implemented 
using the Eq.(\ref{IDS2}). The parameters of the zeroth order solution 
are fixed by requiring $\sigma_{1} = \sigma_{2} =0$ and $\phi_{\rm b}=0$. }
\label{figure6}
\end{figure}
In Fig. \ref{figure6} the back-reaction effects are reported for the case 
of a zeroth-order solution with $\alpha =4$. With 
the dashed line the zeroth-order iterative solution is illustrated.
With the full line the combined effect of the first-order  iteration is illustrated. 
In a radiation-dominated Universe
the evolution equation obeyed by ${\cal F}(\sigma)$ is 
simply given by 
\begin{equation}
\frac{d {\cal F}}{d\sigma} = {\cal F}^2,\qquad {\cal F}^2 = 
\frac{a^{6}}{4} \rho_{\rm r},
\label{RAD}
\end{equation}
where $\rho_{\rm r}$ is a generic radiation fluid with 
$p_{\rm r} =\rho_{\rm r}/3$. As already noticed in connection with 
Eq. (\ref{curv}) the first of the two relations appearing in Eq. (\ref{RAD}) 
implies that the Ricci scalar vanishes in  a radiation-dominated Universe.
The behaviour of the solution for $\sigma \gg \sigma_{\rm b}$,  illustrated 
by the full lines in Fig. \ref{figure6}, corresponds, indeed to the one 
of a radiation-dominated Universe in the $\sigma$ parametrization. In this 
parametrization, in fact ${\cal F} \simeq (\sigma - \sigma_{*})^{-1}$ and 
${\cal F}/a $ is constant.
This non-standard evolution of the scale factor becomes more familiar 
by translating the result from the $\sigma$-parametrization to the 
$\tau$ parametrization. Recalling, in fact, that $a(\sigma)^2 d\sigma = d\tau$, we do 
find that $\tau \simeq (\sigma -\sigma_{*})^{-1}$. This occurrence implies, as expected, that $a(\tau)\simeq \tau$  which is the well known result for a radiation-dominated Universe 
in the conformal time parametrization. 

\begin{figure}
\begin{center}
\begin{tabular}{|c|c|}
      \hline
      \hbox{\epsfxsize = 6.9 cm  \epsffile{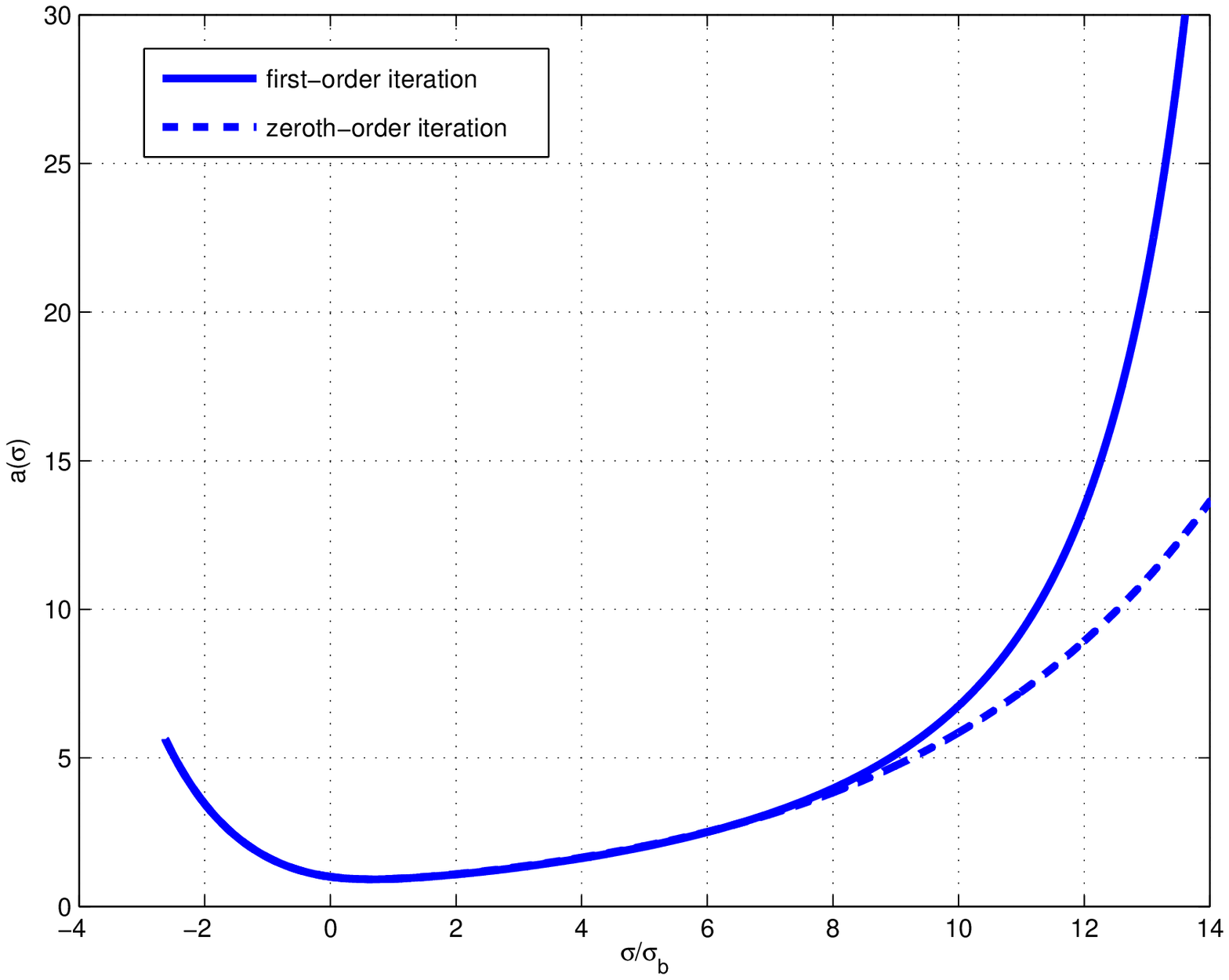}} &
      \hbox{\epsfxsize = 7 cm  \epsffile{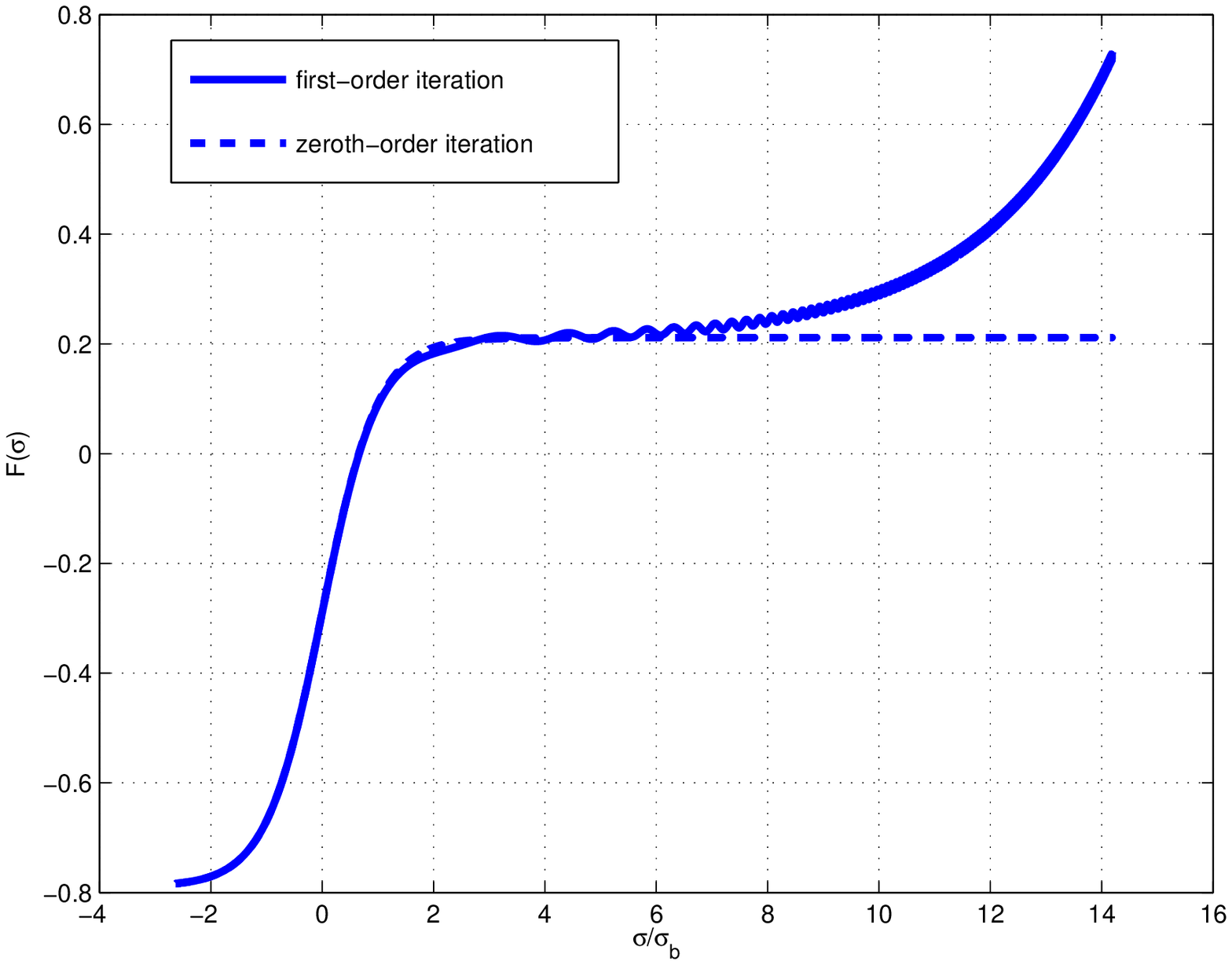}}\\
      \hline
\end{tabular}
\end{center}
\caption{The same quantities reported in Fig. \ref{figure6} but using 
implementing the back-reaction effects through the energy and pressure 
densities stemming from the quadratic corrections to the Einstein tensor
(i.e. Eq. (\ref{IDS3})).}
\label{figure7}
\end{figure}

In Fig. \ref{figure6} the back-reaction effects have been implemented 
by using Eq. (\ref{IDS2}) which comes from Eqs. (\ref{FPrho}) and (\ref{FPp}). In Fig. \ref{figure7} the numerical calculation has been instead performed 
by using Eq. (\ref{IDS3}). By comparing Fig. \ref{figure6} with 
Fig. \ref{figure7} the main visible effect is an enhanced oscillating behaviour of the first-order iteration after the bounce. 
However, the evolution of the scale factor and of ${\cal F}$ is clearly the same.

The iterative method described in the present Section can be also 
applied in the conformal time parametrization. We will omit here 
the technical details but the general idea is to start with the 
asymptotic solution valid for $\tau \ll - \tau_{1}$ (which is, for instance 
the one provided by Eqs. (\ref{anal1}) and (\ref{phanal}) ) and then 
integrate numerically both for the background and for the fluctuations.
The main technical difference between the integration in the 
$\sigma$ and in the $\tau$ parametrization is the following. In the 
$\sigma$ parametrization the analytical solution of Eqs. 
(\ref{Fsol}) and (\ref{xisol}) can be used in the zeroth-order 
iteration. In the $\tau$ parametrization already the zeroth order 
iteration requires numerical treatment. 
Moreover, in the $\tau$ parametrization the evolution 
equations for the mode functions are formally different and 
are  the ones reported in Eq. (\ref{modetau}) 
of the Appendix B.

In Fig. \ref{figure8} the integration of the iterative problem 
is illustrated in the $\tau$ parametrization. With the dashed line 
the zeroth-order solution is reported. With the full line the 
result of the first iteration is illustrated. In the left plot 
of Fig. \ref{figure8} the parameters are chosen to be the same 
ones of Figs. \ref{figure6} and \ref{figure7}. In the right plot the value 
of the gauge coupling at the bounce is chosen to be much smaller 
than $1$. This implies that the scale factor after the bounce is 
much larger than in the case when the gauge coupling is 
of order $1$ at the bounce. This can be understood analytically 
since the asymptotic zeroth-order solution (i.e. Eq. (\ref{anal2}) 
leads to a scale factor whose absolute normalization 
increases as the gauge coupling at the bounce diminishes.
 \begin{figure}
\begin{center}
\begin{tabular}{|c|c|}
      \hline
      \hbox{\epsfxsize = 6.9 cm  \epsffile{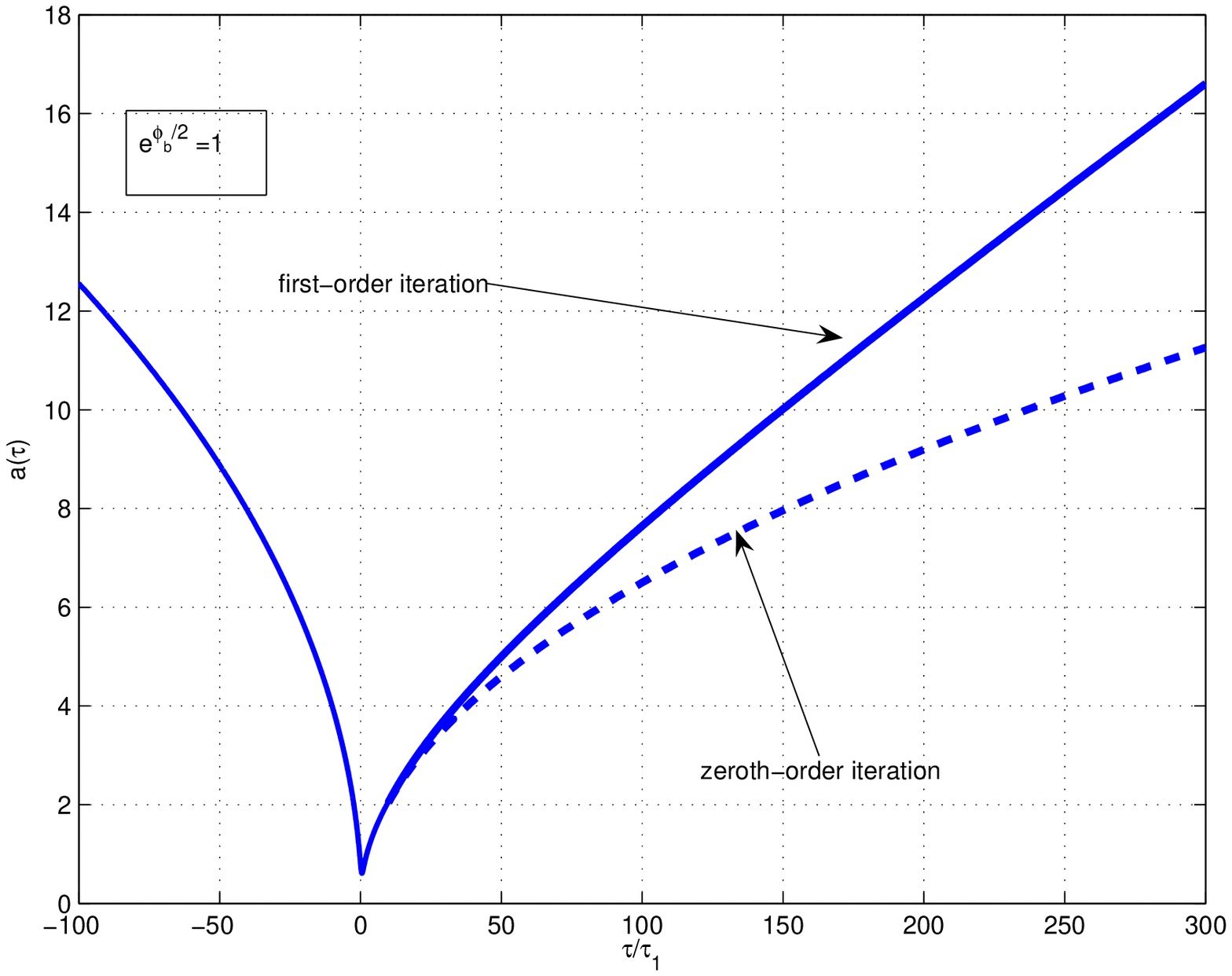}} &
      \hbox{\epsfxsize = 7 cm  \epsffile{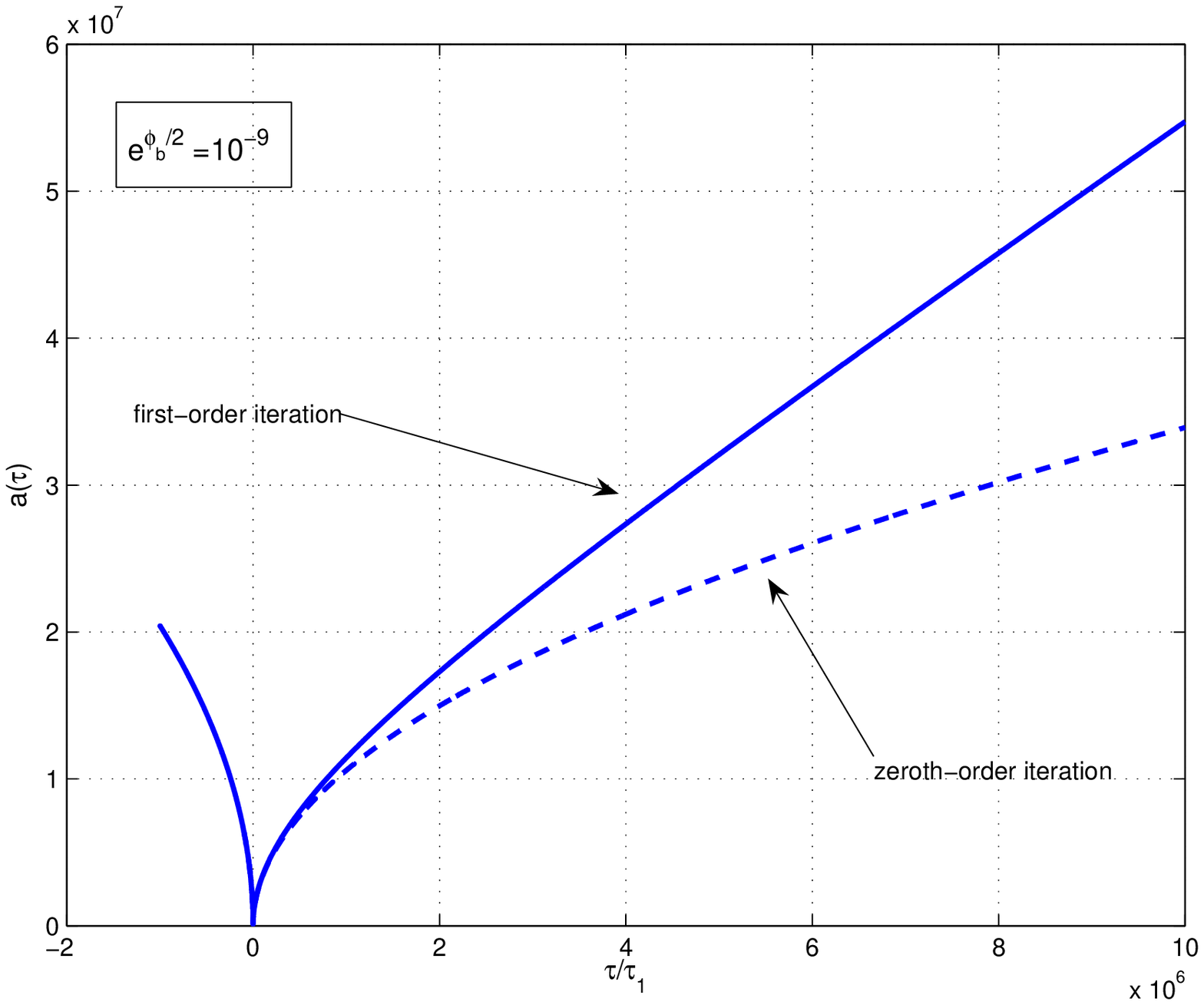}}\\
      \hline
\end{tabular}
\end{center}
\caption{Integration of the iterative problem in the conformal 
time parametrization. In the left plot the parameters of the zeroth-order 
solution coincide (in the limit $\tau \to -\infty$) with the parameters 
chosen for the integrations reported in Figs. \ref{figure6} and 
\ref{figure7} (for $\sigma \to -\infty$). }
\label{figure8}
\end{figure}
It is evident from Fig. \ref{figure8} that the first-order iteration already 
leads to a scale factor that increases linearly in conformal time.
For both plots in Fig. \ref{figure8} the back-reaction 
effects has been parametrized through the analog of Eq. (\ref{IDS2}), 
appropriately translated in the $\tau$ parametrization.
 
\renewcommand{\theequation}{6.\arabic{equation}}
\section{Concluding remarks}
\setcounter{equation}{0}

Dynamical back-reaction of relic gravitons is relevant 
in different contexts, so that the nature of the induced
physical effects may vary. A particularly interesting situation 
is the one of bouncing solutions of pre-big bang type where 
back-reaction effects are known to be important already  from 
qualitative estimates. One of the purposes of the present investigation 
is to propose a theoretical scheme where more quantitative 
predictions could be derived. The 
present findings are then useful for implementing a 
smooth exit from the pre-big bang phase with 
consequent production of radiation. 
The bouncing solutions examined here may also be viewed as 
a practical theoretical laboratory where the methods for the 
analysis of the back-reaction effects of relic gravitons 
may be tested. In this respect, the reported results
suggest that the r\^ole of the produced gravitons 
is crucial for determining the asymptotic state of the solution at late 
times.

The strategy followed in the present analysis relies on the interplay 
between analytical methods and numerical calculations.
We have been interested in situations where the Universe 
undergoes an accelerated contraction before the bounce, while, after 
the bounce, the zeroth-order solution exhibits a decelerated 
expansion. After quantizing the tensor modes of the geometry,
 the initial values of the field operators have been 
set well before the bounce. At that time all the gravitons 
were of short-wavelengths in comparison with the typical 
time scale of the background evolution. Then, the field
operators have been evolved in the Heisenberg representation and 
the mixing coefficients have been accurately determined both 
analytically and numerically. This step allowed  to understand 
what kind of regularization is required for the energy and pressure 
densities of the produced gravitons. The second step has been to 
adopt a consistent ansatz for the gravitational energy-momentum 
pseudo-tensor. Two complementary definitions  of the energy 
and pressure densities of the relic gravitons have then been scrutinized 
and compared both analytically and numerically.
On the basis of the reported results, these two different approaches lead to compatible quantitative results. 

The self-consistent analysis of back-reaction effects leads naturally 
to an integro-differential problem. Consequently, an iterative method for its solution  has been proposed and tested in the 
specific dynamical framework of bouncing solutions 
that arise in the context of the low-energy string effective 
action supplemented by a non-local dilaton potential.
It turns out that the treatment both of the background 
and of the tensor fluctuations becomes rather simple 
by selecting a different parametrization of the time coordinate which 
does not coincide with the usual conformal or cosmic times.
In this time parametrization analytical bouncing solutions 
can be derived in the Einstein frame metric. 

By looking at the structure of the 
low-energy string effective action, it is possible to infer that 
not only relic gravitons but also other massless fields may lead to
similar effects whose analysis is certainly one 
of the possible developments of future studies. In this 
respect, the present investigation dealt with the minimal field 
content of the model with the aim of achieving a reasonably 
accurate description of the underlying physical processes. Thus, some 
of the ideas discussed here will hopefully be useful also for the 
description of dynamical systems with a richer field content.

\newpage
\begin{appendix}
\renewcommand{\theequation}{A.\arabic{equation}}
\setcounter{equation}{0}
\section{Energy-momentum pseudo-tensor}
From Eq. (\ref{pertmet}) we have 
\begin{equation}
\delta_{\rm t}^{(1)} g_{i j} = -  a^2\,\, h_{ij},\qquad h_{i}^{i} = \partial_{i} h^{i}_{j} =0,
\label{MF}
\end{equation}
where, as in the bulk of the paper, the subscript refers to the tensor 
nature of the fluctuation while the superscript denotes the perturbative 
order. Equation (\ref{MF}) implies that 
\begin{equation}
\delta_{\rm t}^{(1)} g^{i j} = \frac{ h^{ij}}{a^2},\qquad
 \delta_{\rm t}^{(2)} g^{i j} = -  \frac{h^{i}_{k} h^{kj}}{a^2}.
\label{metricf}
\end{equation}
Consequently, the  fluctuations of the Christoffel connections to first and 
second order become:
\begin{eqnarray}
&& \delta_{\rm t}^{(1)} \Gamma_{i j}^{0} = \frac{1}{2} ( h_{ij}' + 2 {\cal H} h_{ij}),
\nonumber\\
&& \delta_{\rm t}^{(1)} \Gamma_{i 0}^{j} = \frac{1}{2} {h_{i}^{j}}',
\nonumber\\
&& \delta_{\rm t}^{(1)} \Gamma_{ij}^{k} = 
\frac{1}{2}( \partial_{i} h^{k}_{j} + \partial_{j} h_{i}^{k} - \partial^{k} h_{ij}),
\nonumber\\
&& \delta_{\rm t}^{(2)} \Gamma_{i 0}^{j} = - \frac{1}{2} h^{i k} h_{k j}' ,
\nonumber\\
&& \delta_{\rm t}^{(2)} \Gamma_{ij}^{k} = \frac{1}{2} h^{i \ell} [ \partial_{\ell} h_{j k} -
\partial_{k} h_{j\ell} - \partial_{j} h_{k \ell}],
\label{CONN}
\end{eqnarray}
where, as in the bulk of the paper, the prime denotes a derivation with 
respect to the conformal time coordinate.
The fluctuations of the various components of the Ricci tensor to first 
and second order are then:
\begin{eqnarray}
\delta_{\rm t}^{(1)} R_{ij} &=& \frac{1}{2} [h_{ij}'' + 2 {\cal H} h_{ij}' - \nabla^2 h_{ij}] 
+ ({\cal H}' + 2 {\cal H}^2) h_{ij},
\label{rij1}\\
\delta_{\rm t}^{(2)} R_{00} &=& \frac{1}{4} h_{ij}' {h^{ij}}' - \frac{{\cal H}}{2} 
h_{ij} {h^{ij}}' + \frac{1}{2} h^{ij} \nabla^2 h_{ij},
\label{r002}\\
\delta_{\rm t}^{(2)} R_{ij} &=& \frac{1}{2} h^{k \ell} [ \partial_{k} \partial_{\ell} h_{ij} - 
\partial_{k} \partial_{j} h_{\ell i} - \partial_{k} \partial_{i} h_{j \ell}] 
\nonumber\\
&-& \frac{1}{2} \partial_{j} [ h^{k \ell} ( \partial_{\ell} h_{ik} - \partial_{k} h_{\ell i} - 
\partial_{i} h_{k\ell})] - \frac{{\cal H}}{2} h^{k\ell} h_{k \ell}' \delta_{ij}
\nonumber\\
&+& \frac{{\cal H}}{2} h^{\ell}_{j} h_{\ell i}' + \frac{{\cal H}}{2} h^{\ell}_{i} h_{\ell j}' 
- \frac{1}{4} {h^{k}_{j}}' h_{ik}' - \frac{{\cal H}}{2} {h^{k}_{j}}' h_{ik} - 
\frac{1}{4} {h^{k}_{i}}' h_{kj}' - \frac{{\cal H}}{2} {h^{k}_{i}}' h_{kj}
\nonumber\\
&-& \frac{1}{4} [ \partial_{i} h_{k}^{\ell} + \partial_{k} h^{\ell}_{i} - \partial^{\ell}h_{ik}]
[\partial_{\ell} h^{k}_{j} + \partial_{j} h^{k}_{\ell} - \partial^{k}h_{j \ell}].
\label{rij2}
\end{eqnarray}
The Ricci scalar is zero to first order in the tensor fluctuations, i.e. $\delta_{\rm t}^{(1)} R =0$. This is due to the traceless nature of these fluctuations.
To second-order, however, $\delta_{\rm t}^{(2)} R \neq 0$ and its form is:
\begin{eqnarray}
\delta_{\rm t}^{(2)} R &=& \frac{1}{a^2} \biggl\{ \frac{3}{4} h_{k \ell}' {h^{k\ell}}' + 
{\cal H} h_{k\ell}' h^{k\ell} + \frac{1}{2} h^{k\ell} \nabla^2 h_{k\ell} - 
\frac{1}{4}\partial_{i} h^{k\ell}  \partial^{i} h_{k\ell}\biggr\}
\nonumber\\
&+& \frac{1}{a^2} \biggl\{ - \frac{1}{2} \partial_{i}[ h^{k\ell}(\partial_{\ell} h^{i}_{k} 
- \partial_{k} h_{\ell}^{i} - \partial^{i} h_{k\ell})] 
\nonumber\\
&-& \frac{1}{4}[ \partial_{i} h^{\ell}_{k} \partial_{\ell} h^{k}_{i} - \partial_{i} h^{\ell}_{k}
\partial^{k} h_{i\ell} + \partial_{k} h^{\ell i} \partial_{\ell} h^{k}_{i} - \partial^{\ell} h_{ik}  
\partial^{i} h^{k}_{\ell} + \partial^{\ell} h_{ik} \partial^{k} h_{i \ell}]\biggr\}.
\label{r}
\end{eqnarray}
Recalling the form of the Einstein tensor (see Eq. (\ref{ein}) ),
\begin{equation}
\delta^{(2)}_{\rm t} {\cal G}_{00} = - \ell_{\rm P}^2 {\cal T}_{00}
= \delta_{\rm t}^{(2)} R_{00} - \frac{1}{2} \overline{g}_{00} \delta_{\rm t}^{(2)} R,
\end{equation}
we obtain
\begin{equation}
\ell_{\rm P}^2 {\cal T}_{00} = {\cal H} h_{k\ell}' h^{k\ell} + 
\frac{1}{8} ( h_{k\ell}' {h^{k\ell}}'  + \partial_{i} h_{k\ell} \partial^{i} h^{k\ell}) + 
{\cal D}_{00},
\end{equation} 
where ${\cal D}_{00}$ is a total derivative, i.e. 
\begin{equation}
{\cal D}_{00} = \frac{1}{8} \partial_{\ell}[ \partial_{i} h^{k\ell} h^{i}_{k} - 2 \partial_{k} h^{\ell}_{i} h^{k i}]
\end{equation}

From Eqs. (\ref{rij2}) and (\ref{r}) it is also possible to write:
\begin{eqnarray}
\delta_{\rm t}^{(2)} R_{ij} &=& \frac{1}{4} ( \partial_{k} h^{\ell}_{i} \partial^{k} h_{j\ell} + 
\partial^{\ell} h_{i k} \partial_{\ell} h^{k}_{j}) - 
\frac{1}{4} \partial_{i} h_{k\ell} \partial_{j} h^{k\ell}  
\nonumber\\
&-& \frac{{\cal H}}{2} h^{k\ell} h_{k\ell}' \delta_{ij} + 
\frac{{\cal H}}{2} ( h^{\ell}_{j} h_{\ell i}' + h_{i}^{\ell} h_{\ell j}')
\nonumber\\
&-&\frac{{\cal H}}{2} ( {h^{k}_{j}}' h_{ik} + {h_{i}^{k}}' h_{k j}) - 
\frac{1}{4}({h^{k}_{j}}' h_{ik}' + {h^{k}_{i}}' h_{k j}')
+ {\cal D}_{ij},
\nonumber\\
\delta^{(2)}_{\rm t}R &=& \frac{1}{a^2}\biggl[ \frac{3}{4} h_{k\ell}' {h^{k\ell}}' + 
{\cal H} h_{k\ell}' h^{k\ell} - \frac{3}{4} \partial_{i}h^{k\ell} \partial^{i} h_{k\ell}\biggr]
+ \frac{1}{a^2} {\cal D}_{R}, 
\end{eqnarray}
where ${\cal D}_{ij}$ and ${\cal D}_{R}$ are further total derivative
\begin{eqnarray}
{\cal D}_{ij} &=& \frac{1}{2} \partial_{k} [ h^{k\ell}  ( \partial_{\ell} h_{ij} - 
\partial_{j} h_{i\ell} - \partial_{i} h_{j\ell})]
- \frac{1}{2} \partial_{j} [ h^{k\ell} (\partial_{\ell} h_{ik} - \partial_{k} h_{\ell i} - 
\partial_{i} h_{k\ell})] 
\nonumber\\
&-& \frac{1}{4} \partial_{\ell}[ \partial_{k} h^{\ell}_{i} h^{k}_{j} + 
h_{i k} \partial^{k} h_{j}^{\ell}].
\nonumber\\
{\cal D}_{R} &=& \partial_{i}[ h^{k\ell} \partial^{i} h_{k\ell}] + 
\frac{1}{4}\partial_{\ell} [ \partial_{i} h^{k\ell} h^{i}_{k} 
- 2 \partial_{k} h^{\ell}_{i} h^{k i}],
\end{eqnarray}
Therefore, up to total derivatives, the following result holds:
\begin{equation}
\ell_{\rm P}^2 \,\,{\cal T}_{00} = {\cal H} h_{k\ell}' h^{k\ell} + 
\frac{1}{8} ( h_{k\ell}' {h^{k\ell}}'  + \partial_{i} h_{k\ell} \partial^{i} h^{k\ell}),
\end{equation} 
and
\begin{equation}
\ell_{\rm P}^2\,\, {\cal T}_{ij}  = \frac{3}{8}\delta_{ij}[ \partial_{m} h_{k\ell} \partial^{m} h^{k\ell}
- h_{k\ell}' {h^{k\ell}}'] + \frac{1}{2} {h^{k}_{j}}' h_{ik}' + 
\frac{1}{4} \partial_{i} h_{k\ell} \partial_{j} h^{k\ell} - 
\frac{1}{2} \partial_{k} h_{i}^{\ell} \partial^{k} h_{\ell j}.
\end{equation}
To pass from doubly covariant indices to mixed ones, it is useful to recall that, to second 
order,
\begin{equation}
\delta^{(2)} {\cal G}_{\mu}^{\nu} = \delta^{(2)} [ g^{\nu\alpha} {\cal G}_{\mu\alpha}]
= \delta^{(2)} g^{\nu\alpha} \overline{{\cal G}}_{\mu\alpha} + 
\overline{g}^{\nu\alpha} \delta^{(2)} {\cal G}_{\mu\alpha} + 
\delta^{(1)} g^{\nu\alpha} \delta^{(1)} {\cal G}_{\mu\alpha}.
\end{equation}
By looking at the form of the specific terms arising in the previous equation it is clear that
${\cal T}_{0}^{0} = \overline{g}^{00} {\cal T}_{00}$ and that 
${\cal T}_{i}^{j} = \overline{g}^{jk} {\cal T}_{ki}$.
The expressions for ${\cal T}_{0}^{0}$ and ${\cal T}_{i}^{j}$ are the ones 
reported in Eqs. (\ref{PSsec1}) and (\ref{PSsec2}) (see also Eqs. (\ref{TRsec}) 
and (\ref{ANISsec})). These expressions coincide with the ones 
obtained, for instance, in \cite{abramo1,abramo2} and are also 
consistent with the ones of \cite{isacson1,isacson2}.

\renewcommand{\theequation}{B.\arabic{equation}}
\setcounter{equation}{0}
\section{Analytical estimates of the effective barotropic indices}

Since the Hamiltonian defined in Eq. (\ref{H1t}) 
is time-dependent, it is always possible to perform 
time-dependent canonical transformations, leading to a different 
form of the Hamiltonian that will be classically equivalent to (\ref{H1t}). 
This procedure correspond to drop total time derivatives from the 
corresponding classical action \cite{mgtp}.

From the action (\ref{TA2}), defining the new field $\mu = a h$ 
and dropping a total time derivative the following Hamiltonian 
can be obtained in the $\tau$-parametrization
\begin{equation}
\tilde{H}_{\rm gw}(\tau) = \frac{1}{2}\int d^{3} x \biggl[ \pi^2 + 
(\partial_{i} \mu)^2 - ( {\cal H}^2 + {\cal H}')  \mu^2\biggr],
\label{H2t}
\end{equation}
where $\pi = \mu'$.
This form of the Hamiltonian is particularly convenient for the studying 
the time evolution of the field operators in the conformal time 
parametrization. 
By appropriate Fourier transforms of the field operators (see Eq. (\ref{oper})), the analog of Eqs. (\ref{Heis1a}) and (\ref{Heis1b}) are 
\begin{eqnarray}
\hat{\mu}_{\vk}(\tau) &=& \hat{b}_{\vk}(\tau_{\rm i}) f_{k}(\tau) 
+ \hat{b}^{\dagger}_{-\vk}(\tau_{\rm i})
f_{k}^{*}(\tau),
\nonumber\\
\hat{\pi}_{\vk}(\tau) &=& \hat{b}_{\vk}(\tau_{\rm i}) g_{k}(\tau) + 
\hat{b}^{\dagger}_{-\vk}(\tau_{\rm i}) 
g_{k}^{*}(\tau),
\end{eqnarray}
where the new mode functions now obey 
\begin{equation}
f_{k}'' + \biggl[ k^2 - \frac{a''}{a} \biggr] f_{k}=0,\qquad g_{k} = f_{k}'.
\label{modmod}
\end{equation}
In the limit $\tau_{\rm i}\to -\infty$, the operators $\hat{b}_{\vk}(\tau_{\rm i})$ 
and $\hat{b}_{-\vk}(\tau_{\rm i})$ annihilate the same initial vacuum state 
as the one annihilated by $\hat{a}_{\vk}(\sigma_{\rm i})$ 
and $\hat{a}_{-\vk}(\sigma_{\rm i})$ in the limit $\sigma_{\rm i}\to -\infty$.

In this framework all the calculations discussed in the 
$\sigma$-parametrization can be reformulated. 
The expectation values of the energy 
and pressure density can be computed.  For instance from Eqs. (\ref{PSsec1}) and (\ref{PSsec2}) we have \footnote{In the bulk of the paper, for practical reasons, we dropped 
the expectation values for the averaged quantities. In this appendix, to make clear the 
computational procedures, the expectation values will be restored.}
\begin{eqnarray}
&&\langle \hat{\rho}_{\rm gw} \rangle = \frac{1}{a^4}  \int\frac{d^{3} k}{(2\pi)^3}
\{ ( k^2 -7  {\cal H}^2) |f_{k}(\tau)|^2 + 
|g_{k}(\tau)|^2 + 3 {\cal H} [ f_{k}(\tau) g_{k}^{*}(\tau) + f_{k}^{*}(\tau) g_{k}(\tau)]\},
\label{rhotau}\\
&& \langle \hat{p}_{\rm gw} \rangle = \frac{1}{ 3 a^4}  \int\frac{d^{3} k}{(2\pi)^3}\{ |f_{k}(\tau)|^2 ( 7 k^2 - 5{\cal H}^2 ) - 
5 |g_{k}(\tau)|^2 
\nonumber\\
&& + 5 {\cal H} [f_{k}^{*}(\tau) g_{k}(\tau) + f_{k}(\tau) g_{k}^{*}(\tau)]\}.
\label{ptau}\\
&& \langle \hat{{\cal P}}_{\rm gw} \rangle= \langle \hat{p}_{\rm gw} \rangle
+ 
\frac{4 ( {\cal H}^2 - {\cal H}')}{3 \,{\cal H} \,a^4} \int \frac{d^{3} k}{ (2\pi)^3} 
[ f_{k}(\tau) g_{k}^{*}(\tau) + f_{k}^{*}(\tau) g_{k}(\tau)].
\label{Ptau}
\end{eqnarray}
In similar terms, the expectation values of the energy and pressure 
densities can be obtained within the approach of Refs. \cite{ford1,ford2}.

Consider now the case of a transition from a de Sitter stage 
of expansion to a radiation-dominated stage of expansion, i.e. Eqs. (\ref{dsa}) and (\ref{rada}). For $\tau \leq -\tau_1$, 
\begin{equation}
f_{k}(\tau) = \frac{1}{\sqrt{2 k}} \biggl( 1 - \frac{i}{x}\biggr)\,\, e^{- i x},\qquad
 g_{k}(\tau) = i \sqrt{\frac{k}{2}} 
\biggl[ \frac{1}{x^2} - 1 + \frac{i}{x}\biggr]\,\,e^{- i x},
\label{modf}
\end{equation}
where $ x = k\tau$.
For $\tau >- \tau_{1}$ the canonical fields can be expressed as 
\begin{equation}
\hat{\mu}_{\vk}(\tau) = \hat{b}_{\vk} F_{k}(\tau) + \hat{b}_{-\vk}^{\dagger}F_{k}^{*}(\tau), \qquad
\hat{\pi}_{\vk}(\tau) = \hat{b}_{\vk} G_{k}(\tau) + \hat{b}_{-\vk}^{\dagger}G_{k}^{*}(\tau),
\label{secsol2}
\end{equation}
where the mode functions are now 
\begin{equation}
F_{k}(\tau) = \frac{1}{\sqrt{2 k} } e^{ - i  (x + 2 x_1)},\qquad
  G_{k}(\tau) = -i \sqrt{\frac{k}{2}} e^{- i ( x + 2 x_1)}.
\label{modr3}
\end{equation}
Since the field operators must be continuous we have  
\begin{eqnarray}
f_{k}(-\tau_1) = c_{+}(k) F_{k}(-\tau_1) + c_{-}(k) F_{k}^{*}(-\tau_1),
\nonumber\\
\tilde{g}_{k}(-\tau_1) = c_{+}(k) G_{k}(-\tau_1) + c_{-}(k) G_{k}^{*}(-\tau_1).
\end{eqnarray}
so that 
$c_{+}(k)$ and $c_{-}(k)$ can be determined as
\begin{equation}
 c_{+}(k) = \frac{e^{2 i x_1}( 2 x_1^2 -1 + 2 i x_1)}{2 x_1^2},
\qquad
c_{-}(k) =\frac{1}{2 x_1^2},
\label{AB2}
\end{equation}
where $x_{1} = k\tau_{1}$.
It can be verified immediately that $|c_{+}(k)|^2 - |c_{-}(k)|^2=1$.

It is now instructive to consider the behaviour of the effective barotropic indices in the 
two regimes $\tau < -\tau_{1}$ and $\tau > -\tau_{1}$. Furthermore, in each regime, 
we shall be interested in the relative weight of the short and long wavelength modes. This analysis is useful for the comparison with the case of bouncing models.
Since during the de Sitter phase, ${\cal H}' = {\cal H}^2$,  Eqs. (\ref{ptau}) 
and (\ref{Ptau}) imply that 
$\langle \hat{p}_{\rm gw} \rangle = \langle \hat{{\cal P}}_{\rm gw}\rangle$. Furthermore, 
if we will have that $3\langle \rho_{\rm gw} \rangle = \langle p_{\rm gw} \rangle$.
During the radiation-dominated phase, i.e. $\tau > - \tau_{1}$, the 
short wavelength modes also obey a radiative equation of state and  the energy 
and pressure densities explicitly depend upon the mixing coefficients, i.e. 
\begin{eqnarray}
&&\langle \hat{\rho}_{\rm gw} \rangle = \frac{1}{a^4} 
\int \frac{k d^{3} k}{(2\pi)^3} [ 2 |c_{-}(k)|^2 + 1],
\nonumber\\
&& \langle \hat{{\cal P}}_{\rm gw} \rangle 
=\langle p_{\rm gw} \rangle = \frac{1}{3 a^4} 
\int \frac{k d^{3} k}{(2\pi)^3} [ 2 |c_{-}(k)|^2 + 1],
\end{eqnarray}
where, for simplicity, the vacuum contribution has not been subtracted.
Notice that, now (i.e. in the radiation dominated epoch and for $k \tau \gg 1$)  , the  equality of $\langle \hat{{\cal P}}_{\rm gw} \rangle$ and 
$\langle \hat{p}_{\rm gw} \rangle$, is a consequence 
of the explicit form of the mode functions in this regime, i.e.
$F_{k} G_{k}^{*} + F_{k}^{*} G_{k} =0$.

In the long wavelength limit, i.e. $k \tau \ll 1$ and during the de Sitter phase 
we obtain, instead, the same expressions reported in Eqs. (\ref{exdsa1})  and (\ref{exdsa2}).
Consider finally the last case, i.e. long wavelength modes during the radiation
dominated epoch. 
After making use of the explicit expressions 
of the mixing coefficients we will have, for the averaged energy density,
\begin{eqnarray}
\langle \hat{\rho}_{\rm gw} \rangle &=& \frac{1}{a^4} \int  \frac{ k\,\, d^{3} k}{(2\pi)^3} 
\biggl\{ \frac{1}{2} \biggl[ 2 - \frac{7}{(x + 2 x_1)^2}\biggr] ( 1 + \frac{1}{ 2 x_{1}^4}\biggr)
\nonumber\\
&-& \sin{ 2 (x + x_{1})} \biggl[  \frac{ 3 ( 2 x_{1}^2 -1)}{2 (x + 2 x_1)x_{1}^4} + 
\frac{7}{ 2 (x+ 2 x_1)^2 x_{1}^{3}} \biggr] 
\nonumber\\
&+& \cos{ 2 (x + x_1)} \biggl[ \frac{3}{(x + 2 x_1) x_{1}^3} - 
\frac{7 ( 2 x_{1}^2 -1)}{ 4 x_{1}^4 (x+ 2 x_{1})^2} \biggr] \biggr\},
\label{AVrho1}
\end{eqnarray}
Consequently, to lowest order in $|x|\ll 1$ and $|x_1|\ll 1$ the 
 averaged energy density will be 
\begin{equation}
\langle \hat{\rho}_{\rm gw} \rangle = - \frac{5}{6} \frac{1}{a^2} 
\int \frac{d\ln{k}}{2\pi^2 \tau_{1}^4} k^2,
\end{equation}
which is the result reported in Eq. (\ref{eff1}).
Along similar lines the averaged pressure density can be obtained:
\begin{eqnarray}
\langle \hat{p}_{\rm gw} \rangle &=& \frac{1}{3 a^4} \int \frac{k^4 d\ln{k}}{2\pi^2} 
\biggl\{ \frac{1}{2} \biggl[ 2 - \frac{5}{(x + 2 x_1)^2}\biggr] \biggl( 1 + \frac{1}{2 x_{1}^4}\biggr) 
\nonumber\\
&+& \frac{1}{4 x_{1}^4} \biggl( 12 - \frac{5}{(x+ 2 x_1)^2}\biggr) 
[ ( 2x_{1}^2 - 1) \cos{ 2 (x + x_1)} + 2 x_{1} \sin{ 2  (x + x_1) }]
\nonumber\\
&-& \frac{5}{2 (x + 2 x_{1}) x_{1}^4}[ ( 2 x_{1}^2 -1) \sin{2 (x + x_{1}) } - 2 x_{1} 
\cos{ 2(x + x_1) }] \biggr\},
\label{AVp1}\\
 \langle \hat{{\cal P}}_{\rm gw} \rangle - \langle 
\hat{p}_{\rm gw} \rangle
&=& \frac{8}{ 3 a^{4}} \int \frac{d^{3} k}{( 2 \pi)^3} \biggl[
- \frac{( 2 x_{1}^2 - 1)}{2 x_{1}^4} \sin{ 2(x + x_1) } + 
\frac{1}{x_{1}^3} \cos{ 2(x +x_1)} 
\nonumber\\
&-& 
\frac{1}{(x + 2 x_{1})} \biggl( 1 + \frac{1}{2 x_{1}^4}\biggr) - 
\frac{ ( 2 x_{1}^2 -1)}{2 x x_{1}^4} \cos{ 2  (x+ x_1)} 
\nonumber\\
&-& 
\frac{1}{( x + 2 x_{1}) x_{1}^3}  \sin{2 (x + x_1) }\biggr]
\label{AVP1}
\end{eqnarray}
implying, to lowest order  in $|x|\ll 1$ and $|x_1|\ll1 $, 
\begin{equation}
\langle \hat{p}_{\rm gw} \rangle = \frac{7}{6} \frac{1}{a^2} 
\int \frac{d\ln{k}}{2\pi^2 \tau_{1}^4} k^2,\qquad \langle \hat{{\cal P}}_{\rm gw}\rangle -
 \langle \hat{p}_{\rm gw} \rangle = -\frac{8}{9} \frac{1}{a^2} 
\int \frac{d\ln{k}}{2\pi^2 \tau_{1}^4} k^2,
\end{equation}
which are the results reported in Eqs. (\ref{eff2}) and (\ref{eff2a}). 

Instead of using the energy-momentum pseudo-tensor 
derived from the quadratic corrections to the Einstein tensor, we could use the 
energy and pressure densities proposed in \cite{ford1,ford2} 
(see also \cite{sahni}). In this second case the expectation values of the energy 
and pressure density become:
\begin{eqnarray}
 && \langle \hat{\rho}_{\rm gw} \rangle = 
 \frac{1}{a^4} \int \frac{k^4}{2 \pi^2} d\ln{k} \biggl\{ \frac{1}{2} \biggl[ 2 
 + \frac{1}{( x+ 2 x_1)^2} \biggl( 1 + \frac{1}{2 x_{1}^4}\biggr) 
 \nonumber\\
 &&+ 
 \cos{ 2 (x + x_1)} \biggl[ \frac{( 2 x_{1}^2 -1)}{4 x_{1}^4 ( x + 2 x_1)^2} - 
 \frac{1}{x_1^3 ( x + 2 x_{1})}\biggr] 
 \nonumber\\
 && + 
 \sin{2(x + x_1)} \biggl[ \frac{ ( 2 x_{1}^2 -1)}{ 2 x_{1}^4 ( x + 2 x_1)} + \frac{1}{2 x_{1}^3 ( x + 2 x_1)^2}\biggr]\biggr\},
 \label{AVrho2}\\
 && \langle \hat{p}_{\rm gw} \rangle =  \frac{1}{a^4} \int \frac{k^4}{2 \pi^2} d\ln{k} \biggl\{ \frac{1}{2} \biggl( 1 + \frac{1}{2 x_{1}^4}\biggr) \biggl[ \frac{2}{3} + 
 \frac{1}{( x + 2 x_{1})^2} \biggr]
 \nonumber\\
&&  + \cos{2 (x + x_1)} \biggl[ \frac{ 2 x_1^2 -1 }{4 x_{1}^4} \biggl( \frac{1}{( x+ 2 x_1)^2} - \frac{4}{3} \biggr) - \frac{1}{x_{1}^3 ( x + 2 x_1)} \biggr]
\nonumber\\
&& + \sin{2(x + x_1)} \biggl[ \frac{1}{2 x_1^3} \biggl( \frac{1}{( x + 2 x_1)^2} - \frac{4}{3}\biggr) + \frac{ 2 x_1^2 -1}{2 x_{1}^4 ( x + 2 x_1)} \biggr] \biggr\}
\label{AVp2}
\end{eqnarray}
If we now expand the previous expressions for $|x_1| \ll 1$ and 
for $ | x | \ll 1$ we get to the results reported in Eqs. (\ref{eff3}) and (\ref{eff4}).

In the following we are going to cross check some of the results 
obtained in the case of the $\sigma$ parametrization  but working
directly in the conformal time coordinate.  It should be clear that the 
specific parametrization is not crucial to obtain a given result. However, it can happen that, for practical reasons, a given parametrization has to be preferred since it leads either to more transparent or to swifter results. The idea 
will be here  to solve the evolution of the mode functions exactly 
but keeping an approximate form of the background in the post-bounce 
and in the pre-bounce region. In the conformal time 
parametrization the action given in Eq. (\ref{FPdef}) reads 
\begin{equation}
S_{\rm gw} = \frac{1}{2} \int\, d^{3} x\, d\tau\, a^2 [ (\partial_{\tau} \,h)^2 - a^2 (\partial_{m}\,h)^2],
\end{equation}
whose associate Hamiltonian is 
\begin{equation}
H_{\rm gw} = \frac{1}{2} \int\, d^{3}x\, [ \Pi^2 + a^2(\partial_{m} \,h)^2],
\end{equation}
with canonical momentum given as $\Pi = a^2 h'$. Therefore 
in this case the mode functions obey 
\begin{equation}
\tilde{f}_{k}'' + 2 {\cal H} \tilde{f}_{k}' + k^2 \tilde{f}_{k} =0,\qquad 
\tilde{g}_{k} = a^2 \tilde{f}_{k}'.
\label{modetau}
\end{equation}
Consider, for instance, the case given in 
Eqs. (\ref{anal1}), (\ref{anal2}) and (\ref{phanal}). The form of the 
mode functions before the bounce (i.e. for $\tau < - \tau_{1}$) will then be 
\begin{eqnarray}
\tilde{f}_{k}(\tau) &=& \frac{\sqrt{\tau_{1} \pi}}{\sqrt{2} a_{-}} e^{ i\pi/4} H^{(1)}_{0}(-x),
\nonumber\\
\tilde{g}_{k}(\tau) &=&  - a^{2}(\tau) k \sqrt{\tau_1} e^{i \pi/4} \sqrt{\frac{\pi}{2}} 
H_{1}^{(1)}( - x), 
\end{eqnarray}
where, as in the previous example, $x = k\tau$; $H^{1}_{\nu}(z)$ and 
$H_{\nu}^{(2)}(z)$ will denote, in the following,  the Hankel functions of
index $\nu$ and argument $z$ \cite{abr,tric}.  
For $\tau > \tau_{1}$, recalling Eq. (\ref{anal2}), 
the evolution of the mode functions can be written as:
\begin{eqnarray}
&& \tilde{F}_{k}(\tau) = \frac{\sqrt{\pi}}{\sqrt{2}} \frac{\sqrt{\tau_{1}}}{a_{+}} \biggl[c_{+}(k)\,\,
e^{i\pi/4} \,\, H_{0}^{(1)}(x) + c_{-}(k)\,\,
e^{-i\pi/4} \,\, H_{0}^{(2)}(x)\biggr],
\nonumber\\
&& \tilde{G}_{k}(\tau) = - a^2(\tau)\frac{ k\sqrt{\tau_1}}{a_{+}}
 \sqrt{\frac{\pi}{2}}  \biggl[c_{+}(k)\,\,
e^{i\pi/4} \,\, H_{1}^{(1)}(x) + c_{-}(k)\,\,
e^{-i\pi/4} \,\, H_{1}^{(2)}(x)\biggr].
\end{eqnarray}
It is now instructive to pretend that the analytical form of the mode functions is unavailable 
in the intermediate region, i.e. $-\tau_{1} <\tau < \tau_{1}$. Therefore 
let us try to excise a thin region around the bounce and compute 
the mixing coefficients by assuming
\begin{equation}
\tilde{f}_{k}(-\tau_1) = \tilde{F}_{k}(\tau_{1}),\qquad
\tilde{g}_{k}(-\tau_1) = \tilde{G}_{k}(\tau_{1}).
\end{equation}
This allows to determine 
\begin{eqnarray}
&& c_{+}(k) = - i\,\frac{\pi}{4} x_{1} [ H_{1}^{(2)}(x_1) H_{0}^{(1)}(x_1) + H_{0}^{(2)}(x_1) H_{1}^{(1)}(x_1)],
\nonumber\\
&& c_{-}(k) = - \frac{\pi}{2} x_1 H_{0}^{(1)}(x_1) H_{1}^{(1)}(x_1),
\end{eqnarray}
Notice that from the previous formulas it is possible to obtain, after 
some trivial algebra, that $|c_{+}(k)|^2 - |c_{-}(k)|^2 =1$.
Recalling now that \cite{abr,tric} 
\begin{equation}
 H_{0}^{(1)}(x_1) \sim i \frac{2}{\pi} \ln{x_1},\qquad
H_{1}^{(1)}(x_1) \sim - \frac{2 i}{\pi x_1}
\end{equation}
we  find that for $x_{1} \ll 1$ 
\begin{equation}
c_{-} (k) \sim - \frac{2}{\pi} \ln{x_1}
\end{equation}
Now this result is (qualitatively) similar to the one obtained, by a totally 
different approach, in Eqs. (\ref{CM}), (\ref{CP}) and (\ref{cpmanal}). By doing 
a more careful comparison it is possible to show that the method 
discussed in  Section 3 of the present paper is the most accurate, also 
from a conceptual point of view. However, the path followed in the present appendix is able to reproduce the gross features of the mixing and, therefore, 
can be used for semi-quantitative estimates. 

It is interesting to notice that, within this approach, the results 
of the effective barotropic indices (numerically obtained in the text) 
can be checked. For instance, in the approach of \cite{ford1,ford2} 
the energy and pressure densities before the bounce become, in the 
present example 
\begin{eqnarray}
\langle \hat{\rho}_{\rm gw} \rangle &=& \frac{1}{a^4} \int \frac{k^4}{2 \pi^2}
\, d\ln{k}
\biggl\{ 
\frac{\pi}{4} (-x) [ J_{1}(-x)^2 + Y_{1}(-x)^2 + J_{0}(-x)^2 + Y_{0}(-x)^2]\biggr\},
\nonumber\\
\langle \hat{p}_{\rm gw} \rangle &=& \frac{1}{a^4} \int \frac{k^4}{2 \pi^2}
\, d\ln{k}
\biggl\{ \frac{\pi}{4} (-x)[ J_{1}(-x)^2 + Y_{1}(-x)^2]
\nonumber\\
&& - \frac{\pi}{12} (-x) [J_{0}(-x)^2 + Y_{0}(-x)^2]\biggr\},
\label{prev}
\end{eqnarray}
where $J_{\nu}(z)$ and $Y_{\nu}(z)$ are the Bessel 
functions of order $\nu$ and argument $z$ \cite{abr,tric}.
In the small argument limit, from Eqs. (\ref{prev}) we do get:
\begin{equation}
\langle \hat{p}_{\rm gw} \rangle \simeq 
\langle \hat{\rho}_{\rm gw} \rangle = \frac{1}{a^4} \int \frac{k^4}{2 \pi^2} 
\, d\ln{ k} \biggl( - \frac{1}{\pi x}\biggr),
\label{prev2}
\end{equation}
which is one of the results discussed in Section 4.
Both $\langle \rho_{\rm gw} \rangle $ and 
$ \langle p_{\rm gw} \rangle$ scale as $a^{-6}$ 
(since $a(\tau) \sim \sqrt{-\tau}$)as it is 
consistent with the equation of state 
$\langle \hat{\rho}_{\rm gw} \rangle \simeq\langle \hat{p}_{\rm gw} \rangle$.

The same analysis can be preformed when the energy-momentum 
pseudo-tensor is obtained from the quadratic corrections to the 
Einstein tensor. The result is 
\begin{eqnarray}
\langle \hat{\rho}_{\rm gw} \rangle &=& 
\frac{1}{a^4} \int \frac{k^4}{2 \pi^2} \,\, d\ln{k} \biggl\{ 
\pi [ J_{0}( -x) J_{1}( -x) + Y_{0}(-x) Y_{1}(-x)] 
\nonumber\\
&-&
\frac{\pi}{4} [ J_{0}(-x)^2 + J_{1}(-x)^2 + Y_{0}(-x)^2 +Y_{1}(-x)^2] \biggr\},
\nonumber\\
\langle \hat{p}_{\rm gw} \rangle &=&  \frac{1}{a^4} \int \frac{k^4}{2 \pi^2} \,\, d\ln{k} \biggl\{ \pi [ J_{0}(-x) J_{1}(-x) + Y_{0}(-x) Y_{1}(-x) ] 
\nonumber\\
&-& 
\frac{\pi}{4} x[ J_{0}(-x)^2 + Y_{0}(-x)^2 + J_{1}(-x)^2 + Y_{1}(-x)^2]\biggr\},
\nonumber\\
\langle \hat{{\cal P}}_{\rm gw} \rangle - \langle \hat{p}_{\rm gw} \rangle &=&  
\frac{1}{a^4} \int \frac{k^4}{2 \pi^2} \,\, d\ln{k} 
4\pi  [ J_{0}(-x) J_{1}(-x) +Y_{0}(-x) Y_{1}(-x) ].
\end{eqnarray}
Let us now expand the previous expressions for $|x|\ll 1$.
The result is:
\begin{eqnarray} 
\langle \hat{\rho}_{\rm gw} \rangle &=& \frac{1}{a^4} \int \frac{k^4}{ 2 \pi^2}\, 
d\ln{k} \biggl[ - \frac{1}{\pi x} (  1 - 4 \gamma + 4 \ln{2} - 4 \ln{(-x)} )\biggr],
\nonumber\\
\langle \hat{p}_{\rm gw} \rangle &=& \frac{1}{a^4} \int \frac{k^4}{ 2 \pi^2}\, 
d\ln{k} \frac{5}{ 3 \pi x},
\nonumber\\
\langle \hat{{\cal P}}_{\rm gw} \rangle - \langle \hat{p}_{\rm gw} \rangle &=& 
\frac{1}{a^4} \int \frac{k^4}{ 2 \pi^2}\, 
d\ln{k} \frac{2}{  \pi x} [ 2 \gamma - 2 \ln{2} + 2 \ln{(-x)}].
\label{approx}
\end{eqnarray}
where $ \gamma$ is the Euler-Mascheroni constant \cite{abr,tric}.  From Eqs. (\ref{approx}) it is possible to obtain that in this parametrization of the energy-momentum pseudo-tensor, indeed, logarithmic corrections to the effective 
barotropic index arise, as pointed out in Section 4.  
From the examples reported here it is clear that also the conformal time 
parametrization can be employed for the self-consistent analysis 
of back-reaction effects. Indeed, in Section 5, explicit numerical integrations have 
been reported also in the $\tau$ parametrization. It is however true that, in the 
context of bouncing models, the $\sigma$ coordinate allows to obtain 
the wanted results much faster both analytically and numerically.
\end{appendix}

\end{document}